\documentclass[12pt]{article}
\usepackage{amsmath,latexsym,mathrsfs}
\usepackage{amsfonts} 
\usepackage{amssymb}  
\usepackage{multirow}
\usepackage{colortbl}
\usepackage{graphicx}

\definecolor{gray80}{gray}{0.8}
\newcommand{\ghline}{\arrayrulecolor{gray80}\hline\arrayrulecolor{black}}
\newcommand{\gcline}[1]{\arrayrulecolor{gray80}\cline{#1}\arrayrulecolor{black}}

\DeclareSymbolFont{boldletters}{OML}{cmm} {b}{it}
\DeclareSymbolFontAlphabet{\mathbit}{boldletters}
\DeclareMathSymbol{\alpha}{\mathalpha}{letters}{"0B}
\DeclareMathSymbol{\beta}{\mathalpha}{letters}{"0C}
\DeclareMathSymbol{\gamma}{\mathalpha}{letters}{"0D}
\DeclareMathSymbol{\delta}{\mathalpha}{letters}{"0E}
\DeclareMathSymbol{\epsilon}{\mathalpha}{letters}{"0F}
\DeclareMathSymbol{\zeta}{\mathalpha}{letters}{"10}
\DeclareMathSymbol{\eta}{\mathalpha}{letters}{"11}
\DeclareMathSymbol{\theta}{\mathalpha}{letters}{"12}
\DeclareMathSymbol{\iota}{\mathalpha}{letters}{"13}
\DeclareMathSymbol{\kappa}{\mathalpha}{letters}{"14}
\DeclareMathSymbol{\lambda}{\mathalpha}{letters}{"15}
\DeclareMathSymbol{\mu}{\mathalpha}{letters}{"16}
\DeclareMathSymbol{\nu}{\mathalpha}{letters}{"17}
\DeclareMathSymbol{\xi}{\mathalpha}{letters}{"18}
\DeclareMathSymbol{\pi}{\mathalpha}{letters}{"19}
\DeclareMathSymbol{\rho}{\mathalpha}{letters}{"1A}
\DeclareMathSymbol{\sigma}{\mathalpha}{letters}{"1B}
\DeclareMathSymbol{\tau}{\mathalpha}{letters}{"1C}
\DeclareMathSymbol{\upsilon}{\mathalpha}{letters}{"1D}
\DeclareMathSymbol{\phi}{\mathalpha}{letters}{"1E}
\DeclareMathSymbol{\chi}{\mathalpha}{letters}{"1F}
\DeclareMathSymbol{\psi}{\mathalpha}{letters}{"20}
\DeclareMathSymbol{\omega}{\mathalpha}{letters}{"21}
\DeclareMathSymbol{\varepsilon}{\mathalpha}{letters}{"22}
\DeclareMathSymbol{\vartheta}{\mathalpha}{letters}{"23}
\DeclareMathSymbol{\varpi}{\mathalpha}{letters}{"24}
\DeclareMathSymbol{\varrho}{\mathalpha}{letters}{"25}
\DeclareMathSymbol{\varsigma}{\mathalpha}{letters}{"26}
\DeclareMathSymbol{\varphi}{\mathalpha}{letters}{"27}
\DeclareMathSymbol{\Gamma}{\mathalpha}{letters}{"00}
\DeclareMathSymbol{\Delta}{\mathalpha}{letters}{"01}
\DeclareMathSymbol{\Theta}{\mathalpha}{letters}{"02}
\DeclareMathSymbol{\Lambda}{\mathalpha}{letters}{"03}
\DeclareMathSymbol{\Xi}{\mathalpha}{letters}{"04}
\DeclareMathSymbol{\Pi}{\mathalpha}{letters}{"05}
\DeclareMathSymbol{\Sigma}{\mathalpha}{letters}{"06}
\DeclareMathSymbol{\Upsilon}{\mathalpha}{letters}{"07}
\DeclareMathSymbol{\Phi}{\mathalpha}{letters}{"08}
\DeclareMathSymbol{\Psi}{\mathalpha}{letters}{"09}
\DeclareMathSymbol{\Omega}{\mathalpha}{letters}{"0A}
\newcommand{\mbit}[1]{{\mathbit#1}}

\newcommand{\dbox}{\,\framebox(7,7)[t]{}\,}
 
\begin{document}

\title{
\begin{flushright}
\small{Ehime-th-7}
\end{flushright}
Saddle Points in the Auxiliary Field Method}

\author{Hiroki~Aono and Taro~Kashiwa\thanks{kashiwa@phys.sci.ehime-u.ac.jp}  \\  
Department of Physics, Graduate School of Science and Engineering,\\
 Ehime University, Matsuyama 790-8577, Japan \\
\\}

\date{\today}

\maketitle

\abstract{Investigations are made on the saddle point calculations (SPC) under the auxiliary field method in path integrations. Two different ways of SPC are considered, Method(I) and Method(II), to be checked in an integral representation of the Gamma function, $\Gamma (N)$, as a bosonic example and in a four-fermi type of Grassmann integral where one "fermion mass" $\omega_0$ differs from the other $N$-degenerate species. The recipe of Method(I) seems rather complicated than that of (II) superficially, but the case turns out to be opposite in the actual situation. A general formalism allows us to calculate for $\Gamma (N)$ up to $O \! \left( 1/N^{14} \right)$. It is found that both happen to coincide in the bosonic case but in the fermionic case Method(II) shows a huge deviation in the weak coupling region where $\omega_0 \ll 1$.}

\maketitle\thispagestyle{empty}
\newpage

\section{Introduction}
The auxiliary field method (AFM) is one of the most powerful approximation scheme in path integrations. The recipe for AFM is given as follows: suppose a partition function,
\begin{eqnarray}
Z =   \int d^N \mbit{\sigma }d^N \mbit{\sigma }^* \hspace{1ex}  \exp  \left[  - \omega \left( \mbit{\sigma}^* \cdot \mbit{\sigma} \right)   -   \frac{\lambda^2}{2N} \left(  \mbit{\sigma}^* \cdot \mbit{\sigma} \right)^2  \right]   \ ; \qquad  \left( \mbit{\sigma}^* \cdot \mbit{\sigma} \right)  \equiv \sum_{a=1}^N \sigma_a^* \sigma_a  \ ,   \label{Model}
\end{eqnarray}
where $\sigma_a$'s, $( a=1,\dots , N)$, are fermionic or bosonic degrees. $\omega$ (called a mass) and $\lambda^2$ (a coupling constant) are parameters.

\noindent (i) Introduce the auxiliary fields\cite{rf:KOS, rf:GNKK}, $y$ , (or Hubbard-Stratonovich Field in the solid state physics\cite{rf:FRAD}) by inserting the identity in terms of the Gaussian integral,
\begin{eqnarray}
Z=   \int d^N \mbit{\sigma }d^N \mbit{\sigma }^* \hspace{1ex}   {\mathrm e}^{  - \omega \left( \mbit{\sigma}^* \cdot \mbit{\sigma}  \right)  -   \frac{\lambda^2}{N} \left(  \mbit{\sigma}^* \cdot \mbit{\sigma} \right)^2 } \times    \int_{-\infty}^{\infty}  \! \! \frac{dy}{\sqrt{2 \pi}}   \exp \left[  - \frac{1}{2} \left\{  y +  i \frac{ \lambda }{\sqrt{N}} \left( \mbit{\sigma}^* \cdot \mbit{\sigma} \right)   \right\}^2    \right]   \ , 
\end{eqnarray}
in order to remove the four-body interaction, to obtain
\begin{eqnarray}
Z= \int_{-\infty}^{\infty} \! \!   \frac{dy}{\sqrt{2 \pi}} {\mathrm e}^{-y^2/2}  \int d^N \mbit{\sigma }d^N \mbit{\sigma }^* \hspace{1ex}  \exp \left[ - \left( \omega + i\frac{ \lambda }{\sqrt{N}} y \right)  \left( \mbit{\sigma}^* \cdot \mbit{\sigma}  \right)  \right]     \ .
\end{eqnarray}
(ii) Put $y \mapsto \sqrt{N}y$ and perform the ``Gaussian" type integration of $\sigma_a$  to write
\begin{eqnarray}
E(y)  \equiv  \int d \sigma_a d \sigma_a^*  \hspace{1ex}  \exp \left[ - \left( \omega + i\lambda  y \right)  {\sigma}^*_a \sigma_a  \right]     \ ;  \quad \mbox{for ${}^\forall a$ } \ . \label{E(y)}
\end{eqnarray}
Thus $Z \propto E^N$, yielding
\begin{eqnarray}
 Z = \sqrt{\frac{N}{2 \pi}} \int_{-\infty}^{\infty} \! \! dy \exp \left[ -N \left( \frac{ y^2}{2} - \ln E(y) \right)  \right]   \ . \label{Z'sExpression} 
\end{eqnarray}

\noindent (iii) Write
\begin{eqnarray}
f(y) \equiv \frac{ y^2}{2} - \ln E(y)  \ ,  \label{f(y)Def.}
\end{eqnarray}
and assume $N \mapsto \infty$ to perform a saddle point calculation (SPC): find saddle point(s) $y_0$, satisfying the stability condition,
\begin{eqnarray}
f'(y)\Big|_{y_{{}_0}} = 0 \ ,   \qquad   \mbox{$ f^{(2)}\! \left( y_0 \right) > 0$ ; (stability condition)}    \ ,   
 \label{stabilityCond}
\end{eqnarray}
and expand $f(y)$ around $y_0$, which gives us a power series of $1/N$, called the loop expansion\cite{rf:ZinnKOS}. This is not a convergent but an asymptotic series, of course.

The prescription is simple and straightforward compared to other nonperturbative methods such as the variational\cite{rf:FH} and the optimized perturbation\cite{rf:VOP}. Moreover, the studies in 1- dimensional(= a quantum mechanical) as well as 0-dimensional(= an integration) bosonic\cite{rf:kashiwa} and fermionic\cite{rf:kashiSaka, rf:kashiSaka2} models tell us that we can obtain a fairly accurate result, {\bf even when $N=1$ or small}, from the weak to the strong coupling $\lambda$, by taking higher loops into consideration properly.

Now consider a slightly generalized model;
\begin{eqnarray}
Z^{\mbit \omega } =   \int d^{N+1} \mbit{\sigma }d^{N+1} \mbit{\sigma }^* \hspace{1ex}  \exp  \left[  -  \left( \mbit{\sigma}^* \cdot \mbit{\omega} \cdot \mbit{\sigma} \right)   -   \frac{\lambda^2}{2N} \left(  \mbit{\sigma}^* \cdot \mbit{\sigma} \right)^2  \right]   \ ,  \label{Model1}
\end{eqnarray}
where $(N+1) \times (N+1)$ matrix $\mbit{\omega}$ is given 
\begin{eqnarray}
\mbit{\omega} = \left(
                \begin{array}{cc}
                \omega_0 	&  {\bf 0}^{\rm T} 	\\
               {\bf  0} 	&  \omega {\rm \bf  I}_{\rm N} 
                \end{array}
                \right)   \ , \label{omegaMatrix}
\end{eqnarray}
with ${\rm \bf I}_{\rm N}$ and ${\bf  0}$ being the $N \times N$ unit matrix and the $N$ dimensional zero vector. This corresponds to a field theoretical model of interacting $\sigma_0$ and $\sigma_a \ (a=1,2,\dots N)$ bosons or fermions with the "mass" $\omega_0$ and $\omega $ respectively. Introduce an auxiliary field, $y$, as the above, to find
\begin{eqnarray}
& & Z^{\mbit \omega }= \sqrt{\frac{N}{2 \pi}} \int_{-\infty }^\infty   \! \! dy g(y) \exp \left[ -Nf(y) \right]  \ ,  \label{TargetExp}
\end{eqnarray}
where 
\begin{eqnarray}
g(y) \equiv   \int d \sigma_0 d \sigma_0^*  \hspace{1ex}  \exp \left[ - \left( \omega_0 + i\lambda  y \right)  {\sigma}^*_0 \sigma_0  \right]   \ , \label{g(y)}
\end{eqnarray}
and $f(y)$ has been given by (\ref{f(y)Def.}), with
\begin{eqnarray} 
  E(y )  \equiv   \int d \sigma_a d \sigma_a^*  \hspace{1ex}  \exp \left[ - \left( \omega + i\lambda  y \right)  {\sigma}^*_a \sigma_a  \right]     \ ;  \quad  \mbox{for ${}^\forall a$ }    \ .    
\end{eqnarray}

In view of (\ref{TargetExp}) an issue comes up: there are two ways of performing SPC.
 
\begin{itemize}

 \item Method(I): in\footnote{The integration range of $t$, in (\ref{Method1}) and (\ref{Method2}), need not be specified. Those are given from $-\infty$ to $\infty$ in the final expression under the saddle point method.(See (\ref{Method1Expansion}) to (\ref{M1Expansion2}).)},
\begin{eqnarray}
I_N \equiv \int dt g(t) {\mathrm e}^{-N f(t)} \ , \label{Method1}
\end{eqnarray}
find the saddle point $t_0$ of $f(t)$, $f'(t_0)=0$, then expand $f(t)$ as well as $g(t)$ around $t_0$. Here and hereafter we adopt $t$ instead of $y$ as the integration variable.

 \item Method(II): rewrite (\ref{Method1}) as
\begin{eqnarray}
I_N \equiv \int dt  {\mathrm e}^{-N \tilde{ f}(t)} \ ,  \quad \tilde{ f}(t) \equiv f(t) -\frac{1}{N} \ln g(t)  \  , \label{Method2}  
\end{eqnarray} 
then find the saddle $t_c$ of $\tilde{f}(t)$, $\tilde{f}' \! \left( t_c \right)=0$, and expand $\tilde{f}(t)$ around $t_c$. Finally put $t_c$, given in terms of $1/N$ series, into the expression. 
\end{itemize}

If $N$ becomes large both results would match but, as mentioned above, our interest is to study the validity of AFM when $N$ is small. In this paper, therefore, we study the difference between two methods by considering bosonic and fermionic integrations (0-dimensional field theoretical models). In Sec.\ref{sec.SaddlePoints}, we develop a general formalism of SPC and calculate the asymptotic expansion of the Gamma function as a bosonic model in Sec.\ref{sec:Gamma}. In Sec.\ref{sec:Fermion}, we examine a four-fermi type Grassmann integral and the final Sec.\ref{sec:discussion} is devoted to a discussion.

\section{Saddle Points and the Asymptotic Expansion}\label{sec.SaddlePoints}
In this section, we develop a general formalism of SPC.


\noindent $\bullet$ Method(I): start with the expression (\ref{Method1}) and expand all the integrands around the saddle point, $t_0$, to find
\begin{eqnarray}
 I_N = {\mathrm e}^{-N f_0} \int dt  \left(  \sum_{m=0}^\infty \frac{g^{(m)}_0}{m!}  \left( t-t_0 \right)^m \right)  \exp \left[ - N \frac{f^{(2)}_0}{2} \left( t-t_0 \right)^2 - R_N  \right]  \ , \label{Method1Expansion}  
\end{eqnarray}
where
\begin{eqnarray}
R_N \equiv  N \sum_{n=3}^\infty \frac{f^{(n)}_0}{n!} \left( t-t_0 \right)^n   \ , \label{R_N}
\end{eqnarray}
with
\begin{eqnarray}
f^{(n)}_0 \equiv f^{(n)} \! \left( t_0 \right)  \ ,   \quad  g^{(n)}_0 \equiv g^{(n)} \! \left( t_0 \right)   \ . 
\end{eqnarray}
After checking the stability condition (\ref{stabilityCond}), $f^{(2)}_0 >0$, put $t-t_0 = x/\sqrt{N f^{(2)}_0}$, while assuming $N$ large in (\ref{Method1Expansion}), to find
\begin{eqnarray}
 I_N \approxeq  \frac{{\mathrm e}^{-N f_0}}{\sqrt{N f^{(2)}_0 }}   \sum_{k, m=0}^\infty \frac{(-)^k g^{(m)}_0}{m! k! \left( N f^{(2)}_0 \right)^{m/2} } \int_{-\infty }^{\infty } dx x^m \left( R_N(x) \right)^{k} {\mathrm e}^{- x^2/2} \ , \label{M1Expansion2}
\end{eqnarray}
where (from (\ref{R_N}))
\begin{eqnarray}
R_N(x) \equiv  \frac{1}{\sqrt{N}} \sum_{n=0}^\infty \frac{F_{n+3}}{(n+3)!} \frac{x^{n+3}}{N^{n/2}}  \  ;          \qquad F_n \equiv   \frac{ f_0^{(n)} }{  \left(  f_0^{(2)} \right)^{n/2} }  \ . \label{F_nDefinition}
\end{eqnarray}
(The integration range of $x$ now stretches from $-\infty$ to $\infty$.) Therefore,
\begin{eqnarray}
& & \hspace{-10mm} I_N  \approxeq \frac{{\mathrm e}^{-N f_0}}{\sqrt{N f^{(2)}_0 }}   \sum_{k, m=0}^\infty \frac{(-)^k }{m! k! }  \frac{g^{(m)}_0}{\left( f^{(2)}_0 \right)^{m/2} } \sum_{n_1, n_2,,, n_k = 0}^\infty \left( \frac{1}{N} \right)^{\left(  m + k+\sum_{j=1}^k n_j \right)/2 }   \nonumber \\ 
& & \times \frac{F_{n_1+3}}{(n_1+3)!}  \frac{F_{n_2+3}}{(n_2+3)!} \cdots  \frac{F_{n_k+3}}{(n_k+3)!}  \int_{-\infty }^{\infty } dx \  x^{m + 3k + \sum_{j=1}^k n_j} \  {\mathrm e}^{- x^2/2} \ . \label{I_Nbym,k,n_j}
\end{eqnarray}
Now put 
\begin{eqnarray}
L \equiv  \frac{ m + k+  \sum_{j=1}^k n_j}{2}   \ , \label{LDefinition}
\end{eqnarray}
which must be integers, $L \in \mathbb{Z}$, not half-integers, $L \in \mathbb{Z} + 1/2$; since the power of $x$ in (\ref{I_Nbym,k,n_j}) reads 
\begin{eqnarray}
m + 3k + \sum_{j=1}^k n_j = 2(L + k)  \ , \label{m>0Condition}
\end{eqnarray}
so that the integral vanishes unless $L+k \in \mathbb{Z}$, leaving us
\begin{eqnarray}
\int_{-\infty }^{\infty } dx \  x^{2(L + k)} \  {\mathrm e}^{- x^2/2} = \sqrt{2 \pi } \left( 2(L+k)-1 \right)!!   \ . 
\end{eqnarray}
The expression (\ref{I_Nbym,k,n_j}) then turns out to be
\begin{eqnarray}
 & &   I_N  \approxeq \sqrt{ \frac{2 \pi }{N f^{(2)}_0 }}{\mathrm e}^{-N f_0}  \sum_{L=0}^\infty  \frac{1}{N^L}   \sum_{k=0}^{2L} \frac{(-)^k }{ k! } \left( 2(L+k)-1 \right)!!   \nonumber \\ 
& & \hspace{25ex}  \times  \sum_{ \mbox{ all possible} \ \left\{ n_j \right\} }^{ \sum_{j=1}^k n_j \leq 2L-k  }  \hspace{-2ex}  F(n_1, n_2, \dots , n_k)    \ ,    \label{I_NFinal}  
\end{eqnarray}
where
\begin{eqnarray}
F(n_1, n_2, \dots , n_k) \equiv \frac{1}{(2L -k - \sum_{j=1}^k n_j)!}  \frac{g^{(2L -k - \sum_{j=1}^k n_j)}_0}{\left( f^{(2)}_0 \right)^{(2L -k - \sum_{j=1}^k n_j)/2} }\prod_{j=1}^k  \frac{F_{n_j+3}}{(n_j+3)!}  \ , \label{F Definition}
\end{eqnarray}
is a symmetric function of $n_j$'s. In (\ref{I_NFinal}) the sum should be taken under the condition, 
\begin{eqnarray}
 \sum_{j=1}^k n_j \leq 2L-k \  ;  \label{ConditionalSum<}
\end{eqnarray}
since $m \geq 0$ in (\ref{m>0Condition}). Note if $k=0$, then $ \sum_{j=1}^0 n_j \equiv 0$ and $\prod_{j=1}^0 G(n_j) \equiv 1  $, so that
\begin{eqnarray}
F(n_1, \cdots , n_k) \  \stackrel{k =0}{=} \  \frac{1}{(2L)!}  \frac{g^{(2L)}_0}{\left( f^{(2)}_0 \right)^{L} }   \ .   \label{Cond.k=0} 
\end{eqnarray} 
The conditional sum of $n_j$'s, (\ref{ConditionalSum<}), can be expressed as an alternative form: suppose $Q_\alpha$ of $n_{j}$'s are alike of $A_\alpha $, $n_{j_1}=n_{j_2} = \dots = n_{j_{Q_\alpha} } = A_\alpha$, to write
\begin{eqnarray}
F \! \left(\!  \hspace{4ex}\raisebox{4.5ex}{ $Q_1$}  \hspace{-7.5ex} \overbrace{A_1, \cdots , A_1} ,  \hspace{4ex}\raisebox{4.5ex}{ $Q_2$}  \hspace{-7.5ex} \overbrace{A_2, \cdots , A_2} , \cdots , \ \  \hspace{3.5ex}\raisebox{4ex}{ $Q_P$}  \hspace{-9ex} \overbrace{A_P, \cdots , A_P}\!   \right)  \equiv F \! \left( \left\{ A_1^{Q_1 }  \right\},
\left\{ A_2^{Q_2 }  \right\}, \cdots , \left\{ A_P^{Q_P }  \right\}  \right)  \ . \label{FAQ Definition}
\end{eqnarray}
It is clear that the condition (\ref{ConditionalSum<}) reads as 
\begin{eqnarray}
P \leq k \ ; \qquad \sum_{\alpha=1 }^P Q_\alpha = k \ ,  \quad  \sum_{\alpha=1 }^P Q_\alpha A_\alpha  \left( = \sum_{j=1}^k n_j \right) \leq 2L-k  \ . \label{QA&n_j}
\end{eqnarray}
The multiplicity reads
\begin{eqnarray}
\left(
\begin{array}{c}
 k	\\
 Q_1
\end{array}
\right)   
     \left(
\begin{array}{c}
 k- Q_1	\\
 Q_2
\end{array}
\right)   \dots 
\left(
\begin{array}{c}
 k - \sum_{\alpha=1}^{P-1} Q_\alpha 	\\
 Q_P
\end{array}
\right) = \frac{k!}{Q_1!\cdots Q_P!} \ ,   
\end{eqnarray}
so that (\ref{I_NFinal}) becomes 
\begin{eqnarray}
& & \hspace{-5mm}   I_N  \approxeq  \sqrt{ \frac{2 \pi }{N f^{(2)}_0 }}{\mathrm e}^{-N f_0}  \sum_{L=0}^\infty  \frac{1}{N^L}  \sum_{k=0}^{2L} (-)^k \left( 2(L+k)-1 \right)!!  \nonumber \\ 
& &  \hspace{0mm}  \times  \sum_{\mbox{all possible} \ \{ A_\alpha \} \hfill \atop
\scriptstyle  \sum_{\alpha=1 }^P Q_\alpha= k \  ; \  \left(  P \leq k  \right) }^{ \sum_{\alpha=1 }^P Q_\alpha A_\alpha  \leq 2L-k }  
 \hspace{-2ex} \frac{1}{Q_1!\cdots Q_P!} F \! \left( \left\{ A_1^{Q_1 }  \right\},
\left\{ A_2^{Q_2 }  \right\}, \dots , \left\{ A_P^{Q_P }  \right\}  \right)  \ ,   \label{I_NFinal3}  
\end{eqnarray}
where we have assumed that $A_1 < A_2 < \dots < A_P$.
This is the main formula of the Method(I). Let us classify $I_N$, according to the WKB-approximation \cite{rf:ZinnKOS} under which $1/\hbar $ appears instead of $N$;
\begin{enumerate}
 \item Tree:
\begin{eqnarray}
 \left( I_N  \right)_{\rm tree} \equiv {\mathrm e}^{-N f_0} g_0  \ . \label{TreePart Def}
\end{eqnarray}
 \item $l$-loop: terms up to $l \left( \geq 1 \right)  $ in (\ref{I_NFinal}):
\begin{eqnarray}
 \left( I_N  \right)_{\mbox{$l$-loop}} \equiv {\mathrm e}^{-N f_0} \sqrt{ \frac{2 \pi }{N f^{(2)}_0 }} g_0 \left( 1 + \cdots +  O\! \left( \frac{1}{N^{l-1}} \right)  \right)  \ . \label{L-loop Def}
\end{eqnarray}
\end{enumerate}


\noindent $\bullet $ Method(II): by putting $f(t) \mapsto \tilde{f}(t) \ ; \  g(t) \mapsto 1$ and $t_0 \mapsto t_c$, all expressions in Method(I) can be read as those of Method(II). Write 
\begin{eqnarray}
\tilde{f}^{(n)}_c \equiv \tilde{f}^{(n)} \! \left( t_c \right)  \ ,  
\end{eqnarray}
then (\ref{I_NFinal}) is changed to
\begin{eqnarray}
& & \hspace{0mm}   I_N  \approxeq \sqrt{ \frac{2 \pi }{N \tilde{f}^{(2)}_c }}{\mathrm e}^{-N \tilde{f}_c}  \sum_{L=0}^\infty  \frac{1}{N^L}   \nonumber \\ 
& & \hspace{4mm}  \times  \sum_{k=0}^{2L} \frac{(-)^k }{ k! } \left( 2(L+k)-1 \right)!!  \sum_{\mbox{all possible} \ \{ n_j \} }^{ \sum_{j=1 }^k n_j  = 2L-k }  \hspace{-2ex}\tilde{ F}(n_1, n_2, \dots , n_k)    \ ,    
\label{I_NTildeFinal}  \\
& & \hspace{-5mm} \tilde{ F}(n_1, n_2, \dots , n_k) \equiv \prod_{j=1}^k  \frac{\tilde{F}_{n_j+3}}{(n_j+3)!}  \ ; \qquad \tilde{F}_{n} \equiv \frac{\tilde{f}^{(n)}_c}{ \left( \tilde{f}^{(2)}_c \right)^{n/2} }  \ .  \label{TildeF Definition}
\end{eqnarray}
It should be noted that without $g(t)$ the conditional sum is given {\bf by the equality, $\sum_{j=1}^k n_j= 2L-k$, not by the inequality} (\ref{ConditionalSum<}). Also (\ref{I_NFinal3}) is changed to
\begin{eqnarray}
& & \hspace{-6ex}   I_N  \approxeq \sqrt{ \frac{2 \pi }{N \tilde{f}^{(2)}_c }}{\mathrm e}^{-N \tilde{f}_c}  \sum_{L=0}^\infty   \frac{1}{N^L}   \sum_{k=0}^{2L}(-)^k \left( 2(L+k)-1 \right)!!  \nonumber \\ 
& &  \hspace{5ex}  \times \hspace{0ex}  \sum_{\mbox{all possible} \ \{ A_\alpha \} \hfill \atop
\scriptstyle  \sum_{\alpha=1 }^P Q_\alpha= k \  ; \  \left(  P \leq k  \right) }^{ \sum_{\alpha=1 }^P Q_\alpha A_\alpha  = 2L-k }  \frac{1}{Q_1!\cdots Q_P!} \tilde{F} \! \left( \left\{ A_1^{Q_1 }  \right\},
\left\{ A_2^{Q_2 }  \right\}, \dots , \left\{ A_P^{Q_P }  \right\}  \right)  \ ,   \label{I_NTildeFinal3}  
\end{eqnarray}
where $\tilde{F} \! \left( \left\{ A_1^{Q_1 }  \right\}, \left\{ A_2^{Q_2 }  \right\}, \dots , \left\{ A_P^{Q_P }  \right\}  \right)$ is defined by the expression, (\ref{FAQ Definition}), with the tildes. Here again the conditional sum (\ref{QA&n_j}) becomes simpler, $\sum_{\alpha=1 }^P Q_\alpha A_\alpha   = 2L-k$. Classification in this case is again
\begin{enumerate}
 \item Tree:
\begin{eqnarray}
 \left( I_N  \right)_{\rm tree} \equiv {\mathrm e}^{-N \tilde{f}_c^{(0)}}   \ , \label{TreePart Def Methd2}
\end{eqnarray}
 \item $l$-loop: terms up to $l \left( \geq 1 \right)  $ in (\ref{I_Nbym,k,n_j}) or (\ref{I_NFinal}):
\begin{eqnarray}
 \left( I_N  \right)_{\mbox{$l$-loop}} \equiv {\mathrm e}^{-N \tilde{f}_c^{(0)}} \sqrt{ \frac{2 \pi }{N \tilde{f}^{(2)}_c }}  \left( 1 + \cdots +  O\! \left( \frac{1}{N^{l-1}} \right)  \right)  \ . \label{L-loop Def Method2}
\end{eqnarray}
\end{enumerate}
Note that there are additional powers of $1/N$ hidden in $t_c$. The final task is then {\bf to expand all functions of $t_c$ up to $\left( 1/N \right)^{{l-1}}$}, which however depends on individual models so is relegated to the following sections.

\section{A Bosonic Case: the Gamma Function}\label{sec:Gamma}
As a simple bosonic example, in this section we consider the Gamma function,
\begin{eqnarray}
 \Gamma (N) = \int_{0}^{\infty } dt t^{N-1} {\mathrm e}^{-t} = N^N I_N  \  ;   \quad  
 I_N \equiv \int_{0}^{\infty } dt \left( \frac{1}{t} \right)  {\mathrm e}^{-N \left( t-\ln t \right) } \ , \label{IntegralI_N}
\end{eqnarray}
where we have put $t \mapsto Nt$ in the final expression. Then 
\begin{eqnarray}
\mbox{Method(I)} \ : & & f(t) \equiv t - \ln t \ ;  \quad g(t) \equiv  \frac{1}{t}      \  .  \\
\mbox{Method(II)} \ : & & \tilde{f}(t)  \equiv t - \left( 1-\frac{1}{N} \right) \ln t    \  . \label{Method2f}
\end{eqnarray}
\noindent $\bullet $Method(I): the saddle point, $t_0$ is given by 
\begin{eqnarray}
f'(t_0)= 1 - \frac{1}{t_0} = 0  \Longrightarrow t_0=1  \ , 
\end{eqnarray}
so that
\begin{eqnarray}
f_0 = 1 = g_0 \  , \quad  f^{(n+1)}_0 = - g^{(n)}_0 = - (-)^n n!  \ ; \ \left( n \geq 1 \right) \ , 
\end{eqnarray}
which ensures the stability condition, $f^{(2)}_0 = 1 >0$, and
\begin{eqnarray}
F(n_1, n_2, \dots, n_k) = \prod_{j=1}^k \frac{1}{(n_j+3)}   \ ,
\end{eqnarray}
from (\ref{F Definition}). Therefore from (\ref{I_NFinal}) and (\ref{I_NFinal3}),
\begin{eqnarray}
& & \hspace{-3ex}    I_N \approxeq  \sqrt{ \frac{2 \pi }{N }}{\mathrm e}^{-N}  \sum_{L=0}^\infty  \frac{1}{N^L}  \sum_{k=0}^{2L}   T \! \left( L,k | \leq  2L-k \right)   \ , \label{Gamma'sExpressionWithInequality} 
\end{eqnarray}
where
\begin{eqnarray}
 T \! \left( L,k |\leq 2L-k \right) &  \equiv &   \frac{(-)^k}{k!} \left( 2(L+k) -1\right)!!  \sum_{\mbox{all possible $\{n_j\}$ } }^{\sum_{j=1}^k n_j\leq 2L-k}  \prod_{j=1}^k \frac{1}{(n_j+3)}    \label{T(L,k| < )Def.n_k}  \\ 
&   = & (-)^k  \left( 2(L+k) -1\right)!! \   \sum_{\scriptstyle \mbox{all possible} \ \{ A_\alpha \} \hfill \atop
\scriptstyle  \sum_{\alpha=1 }^P Q_\alpha= k \  ; \  \left(  P \leq k  \right) }^{ \sum_{\alpha=1 }^P Q_\alpha A_\alpha  \leq 2L-k }    \frac{1}{Q_1!\cdots Q_P!}  \nonumber \\ 
& & \hspace{10ex} \times     \frac{1}{\left( A_1 + 3 \right)^{Q_1} \left( A_2 + 3 \right)^{Q_2} \cdots \left( A_P + 3 \right)^{Q_P}} \  . \label{T(L,k| < )Def.Q}
\end{eqnarray}
Here we concentrate on the latter expression (\ref{T(L,k| < )Def.Q}) and realize that most of the terms in the conditional sum, $  \sum_{\alpha =1 }^P Q_\alpha A_\alpha \leq 2L-k$, are canceled, leaving only the terms satisfying the equality $\sum_{\alpha =1 }^P Q_\alpha A_\alpha = 2L-k$ , that is 
\begin{eqnarray}
\sum_{k=0}^{2L}   T \! \left( L,k | \leq 2L-k \right) = \sum_{k=0}^{2L}   T \! \left( L,k | 2L-k \right) \ ,  \label{Equality=InEqality}
\end{eqnarray}
to give
\begin{eqnarray}
 I_N \approxeq  \sqrt{ \frac{2 \pi }{N  }}{\mathrm e}^{-N}  \sum_{L=0}^\infty  \frac{1}{N^L}  \sum_{k=0}^{2L}   T \! \left( L,k | 2L-k \right)  \ , \label{GammaWithEquality}  
\end{eqnarray}
where 
\begin{eqnarray}
& & \hspace{-5ex} T \! \left( L,k | 2L-k \right)   \equiv    \frac{(-)^k}{k!} \left( 2(L+k) -1\right)!!  \sum_{\mbox{all possible $\{n_j\}$ } }^{\sum_{j=1}^k n_j = 2L-k} \prod_{j=1}^k \frac{1}{(n_j+3)}      \label{T(L,k| )Def.n_k}  \\ 
& & \hspace{10ex}  =  (-)^k  \left( 2(L+k) -1\right)!! \  \sum_{\mbox{all possible} \ \{ A_\alpha \} \hfill \atop
\scriptstyle  \sum_{\alpha=1 }^P Q_\alpha= k \  ; \  \left(  P \leq k  \right) }^{ \sum_{\alpha=1 }^P Q_\alpha A_\alpha  = 2L-k }   \frac{1}{Q_1!\cdots Q_P!}  \nonumber \\ 
& & \hspace{15ex} \times     \frac{1}{\left( A_1 + 3 \right)^{Q_1} \left( A_2 + 3 \right)^{Q_2} \cdots \left( A_P + 3 \right)^{Q_P}} \ , \label{T(L,k|  )Def.Q} 
\end{eqnarray}
with the definition when $k=0$, (implying $P=0$), and $L \neq 0$
\begin{eqnarray}
 T \! \left( L, 0 | 2L \right) =  0  \ , 
\end{eqnarray}
since otherwise there is inconsistency in the conditional sum $\sum_{\alpha=1 }^{P=0} Q_\alpha A_\alpha (=0)  $ and $2L (\neq 0)$. The numerical values of $ T \! \left( L,k | 2L-k \right) $ (apart from the factor $\left( 2(L+k) -1\right)!!$) are listed in the appendix \ref{sec:appendixA}. The proof of (\ref{Equality=InEqality}) is rather lengthy then relegated to the appendix \ref{sec:appendixB}.


In view of (\ref{I_NFinal}) and (\ref{F Definition}), the above fact implies that there is no contribution from $g^{(m)}_0 (m \geq 1)$, that is, the equality holds
\begin{eqnarray}
  \int_{0}^{\infty } dt \left( \frac{1}{t} \right)  {\mathrm e}^{-N \left( t-\ln t \right) }\stackrel{1/N } {= } \int_{0}^{\infty } dt  {\mathrm e}^{-N \left( t-\ln t \right) }  \ , \label{no1/Ncontribution}
\end{eqnarray}
under the $1/N$ expansion (Actually it does hold without $1/N$ in this case: see the expression (\ref{FormulaGammaCase}). A detailed discussion is relegated to the appendix \ref{sec:appendixB}.)


In view of (\ref{TreePart Def}) as well as (\ref{IntegralI_N}), 
\begin{eqnarray}
 \left( \Gamma(N) \right)_{\rm tree}=  N^N e^{-N}  \ . 
\end{eqnarray}
And the result of 15-loop is from (\ref{L-loop Def}),
\begin{align}
\left( \Gamma(N) \right)_{\rm 15-loop} & = \left( \sqrt{\frac{2\pi}{N}} N^N e^{-N} \right)\Biggl\{ 1+\frac{1}{12} \frac{1}{N}+\frac{1}{288 }\frac{1}{N^2}   -\frac{139}{51840 }\frac{1}{N^3}-\frac{571}{2488320 }\frac{1}{N^4}   \nonumber \\
&  +\frac{163879}{209018880 }\frac{1}{N^5}+\frac{5246819}{75246796800 }\frac{1}{N^6} - \frac{534703531}{902961561600 }\frac{1}{N^7} \nonumber \\
&-\frac{4483131259}{86684309913600 }\frac{1}{N^8} +\frac{432261921612371}{514904800886784000 } \frac{1}{N^9}\nonumber\\ 
& + \frac{6232523202521089}{86504006548979712000 }\frac{1}{N^{10}}  -\frac{25834629665134204969}{13494625021640835072000 } \frac{1}{N^{11}}   \nonumber\\
& -\frac{1579029138854919086429}{9716130015581401251840000 } \frac{1}{N^{12}} +\frac{746590869962651602203151}{116593560186976815022080000 }\frac{1}{N^{13}}\nonumber\\
& +\frac{1511513601028097903631961}{2798245444487443560529920000 } \frac{1}{N^{14}} + O \! \left( N^{-15} \right)   \Biggr\}   \ . \label{gamma14}
\end{align}
This coincides the results in references\footnote{We have no information how those results are obtained, but the derivations would be from another prescription such as the Watson's Lemma\cite{rf:Copson}}; up to $1/N^4$\cite{rf:Ryz}, $1/N^7$\cite{rf:Iwa} and $1/N^9$\cite{rf:Wolfram}. 
\begin{table}[h]
{\tiny
$$
\begin{array}{|c|cccc|}
\hline\hline
  & N=1 & N=2 & N=5 & N=10 \\
\hline
 \textrm{Exact} & 1 & 1 & 24 & 362880 \\
\hline
 \begin{array}{c} \textrm{ tree}\\ \textrm{ (tree)/Exact}\\ \end{array} &
  \begin{array}{c} 0.36788  \\ 0.36788  \\ \end{array} &
  \begin{array}{c} 0.54134  \\ 0.54134  \\ \end{array} &
  \begin{array}{c} 2.10561\times 10^1 \\ 0.87734  \\ \end{array} &
  \begin{array}{c} 4.53999\times 10^5 \\ 1.25110  \\ \end{array} \\
\ghline
 \begin{array}{c} \textrm{ 1-loop}\\ \textrm{ (1-loop)/Exact}\\ \end{array} &
  \begin{array}{c} 0.92214  \\ 0.92214  \\ \end{array} &
  \begin{array}{c} 0.95950  \\ 0.95950  \\ \end{array} &
  \begin{array}{c} 2.36038\times 10^1 \\ 0.98349  \\ \end{array} &
  \begin{array}{c} 3.59870\times 10^5 \\ 0.99170  \\ \end{array} \\
\ghline
 \begin{array}{c} \textrm{ 2-loop}\\ \textrm{ (2-loop)/Exact}\\ \end{array} &
  \begin{array}{c} 0.99898  \\ 0.99898  \\ \end{array} &
  \begin{array}{c} 0.99948  \\ 0.99948  \\ \end{array} &
  \begin{array}{c} 2.39972\times 10^1 \\ 0.99988  \\ \end{array} &
  \begin{array}{c} 3.62868\times 10^5 \\ 0.99997  \\ \end{array} \\
\ghline
 \begin{array}{c} \textrm{ 3-loop}\\ \textrm{ (3-loop)/Exact}\\ \end{array} &
  \begin{array}{c} 1.00218  \\ 1.00218  \\ \end{array} &
  \begin{array}{c} 1.00031  \\ 1.00031  \\ \end{array} &
  \begin{array}{c} 2.40005\times 10^1 \\ 1.00002  \\ \end{array} &
  \begin{array}{c} 3.62881\times 10^5 \\ 1.00000  \\ \end{array} \\
\ghline
 \begin{array}{c} \textrm{ 4-loop}\\ \textrm{ (4-loop)/Exact}\\ \end{array}&
  \begin{array}{c} 0.99971  \\ 0.99971  \\ \end{array} &
  \begin{array}{c} 0.99999  \\ 0.99999  \\ \end{array} &
  \begin{array}{c} \rowcolor[gray]{0.8}2.40000\times 10^1 \\ \rowcolor[gray]{0.8}1.00000  \\ \end{array} &
  \begin{array}{c} \rowcolor[gray]{0.8}3.62880\times 10^5 \\ \rowcolor[gray]{0.8}1.00000  \\ \end{array} \\
\ghline
 \begin{array}{c} \textrm{ 5-loop}\\ \textrm{ (5-loop)/Exact}\\ \end{array} &
  \begin{array}{c} 0.99950  \\ 0.99950  \\ \end{array} &
  \begin{array}{c} 0.99998  \\ 0.99998  \\ \end{array} &
  \begin{array}{c} 2.40000\times 10^1 \\ 1.00000  \\ \end{array} &
  \begin{array}{c} 3.62880\times 10^5 \\ 1.00000  \\ \end{array} \\
\ghline
 \begin{array}{c} \textrm{ 6-loop}\\ \textrm{ (6-loop)/Exact}\\ \end{array} &
  \begin{array}{c} \rowcolor[gray]{0.8}1.00022  \\ \rowcolor[gray]{0.8}1.00022  \\ \end{array} &
  \begin{array}{c} \rowcolor[gray]{0.8}1.00000  \\ \rowcolor[gray]{0.8}1.00000  \\ \end{array} &
  \begin{array}{c} 2.40000\times 10^1 \\ 1.00000  \\ \end{array} &
  \begin{array}{c} 3.62880\times 10^5 \\ 1.00000  \\ \end{array} \\
\ghline
 \begin{array}{c} \textrm{ 7-loop}\\ \textrm{ (7-loop)/Exact}\\ \end{array} &
  \begin{array}{c} 1.00029  \\ 1.00029  \\ \end{array} &
  \begin{array}{c} 1.00000  \\ 1.00000  \\ \end{array} &
  \begin{array}{c} 2.40000\times 10^1 \\ 1.00000  \\ \end{array} &
  \begin{array}{c} 3.62880\times 10^5 \\ 1.00000  \\ \end{array} \\
\ghline
 \begin{array}{c} \textrm{ 8-loop}\\ \textrm{ (8-loop)/Exact}\\ \end{array} &
  \begin{array}{c} 0.99974  \\ 0.99974  \\ \end{array} &
  \begin{array}{c} 1.00000  \\ 1.00000  \\ \end{array} &
  \begin{array}{c} 2.40000\times 10^1 \\ 1.00000  \\ \end{array} &
  \begin{array}{c} 3.62880\times 10^5 \\ 1.00000  \\ \end{array} \\
\ghline
 \begin{array}{c} \textrm{ 9-loop}\\ \textrm{ (9-loop)/Exact}\\ \end{array} &
  \begin{array}{c} 0.99969  \\ 0.99969  \\ \end{array} &
  \begin{array}{c} 1.00000  \\ 1.00000  \\ \end{array} &
  \begin{array}{c} 2.40000\times 10^1 \\ 1.00000  \\ \end{array} &
  \begin{array}{c} 3.62880\times 10^5 \\ 1.00000  \\ \end{array} \\
\ghline
 \begin{array}{c} \textrm{10-loop}\\ \textrm{ (10-loop)/Exact}\\ \end{array} &
  \begin{array}{c} 1.00047  \\ 1.00047  \\ \end{array} &
  \begin{array}{c} 1.00000  \\ 1.00000  \\ \end{array} &
  \begin{array}{c} 2.40000\times 10^1 \\ 1.00000  \\ \end{array} &
  \begin{array}{c} 3.62880\times 10^5 \\ 1.00000  \\ \end{array} \\
\ghline
 \begin{array}{c} \textrm{11-loop}\\ \textrm{ (11-loop)/Exact}\\ \end{array} &
  \begin{array}{c} 1.00053  \\ 1.00053  \\ \end{array} &
  \begin{array}{c} 1.00000  \\ 1.00000  \\ \end{array} &
  \begin{array}{c} 2.40000\times 10^1 \\ 1.00000  \\ \end{array} &
  \begin{array}{c} 3.62880\times 10^5 \\ 1.00000  \\ \end{array} \\
\ghline
 \begin{array}{c} \textrm{12-loop}\\ \textrm{ (12-loop)/Exact}\\ \end{array} &
  \begin{array}{c} 0.99877  \\ 0.99877  \\ \end{array} &
  \begin{array}{c} 1.00000  \\ 1.00000  \\ \end{array} &
  \begin{array}{c} 2.40000\times 10^1 \\ 1.00000  \\ \end{array} &
  \begin{array}{c} 3.62880\times 10^5 \\ 1.00000  \\ \end{array} \\
\ghline
 \begin{array}{c} \textrm{13-loop}\\ \textrm{ (13-loop)/Exact}\\ \end{array} &
  \begin{array}{c} 0.99862  \\ 0.99862  \\ \end{array} &
  \begin{array}{c} 1.00000  \\ 1.00000  \\ \end{array} &
  \begin{array}{c} 2.40000\times 10^1 \\ 1.00000  \\ \end{array} &
  \begin{array}{c} 3.62880\times 10^5 \\ 1.00000  \\ \end{array} \\
\ghline
 \begin{array}{c} \textrm{14-loop}\\ \textrm{ (14-loop)/Exact}\\ \end{array} &
  \begin{array}{c} 1.00452  \\ 1.00452  \\ \end{array} &
  \begin{array}{c} 1.00000  \\ 1.00000  \\ \end{array} &
  \begin{array}{c} 2.40000\times 10^1 \\ 1.00000  \\ \end{array} &
  \begin{array}{c} 3.62880\times 10^5 \\ 1.00000  \\ \end{array} \\
\ghline
 \begin{array}{c} \textrm{15-loop}\\ \textrm{ (15-loop)/Exact}\\ \end{array} &
  \begin{array}{c} 1.00502  \\ 1.00502  \\ \end{array} &
  \begin{array}{c} 1.00000  \\ 1.00000  \\ \end{array} &
  \begin{array}{c} 2.40000\times 10^1 \\ 1.00000  \\ \end{array} &
  \begin{array}{c} 3.62880\times 10^5 \\ 1.00000  \\ \end{array} \\
\hline\hline
\end{array}
$$

\vspace{-3ex}

\caption{Results of Method(I) in $N=1,2,5$ and $10$ up to $L=14$. Even in $N=1$, 2-loop($L=1$) approximation is sufficiently close to the exact value, however, deviation becomes gradually eminent in the loops larger than 7. The optimized values are shaded in each $N$, whose position shows that the $1/N$ expansion is indeed an asymptotic one.}\label{Tab.Method(I)}
}
\end{table}


\begin{figure}[h]
$$
\includegraphics[width=6cm]{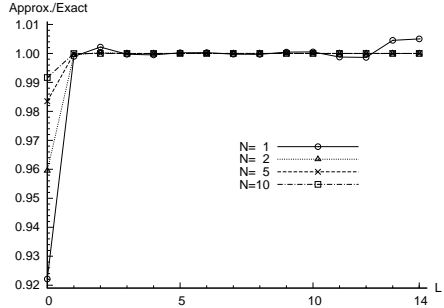} 
$$

\vspace{-3ex}

\caption{The ratio of Approximation to Exact in Method(I): the horizontal axis designates $L$, $0 \leq L \leq 14$, omitting the tree part. The solid line with the circle, the dotted with the triangle, the dashed with the cross, and the dash-dotted with the square designate $N=1, 2, 5$ and 10 respectively. The tail end of the $N=1$ line deviates from the unity, implying the asymptotic character of the $1/N$ expansion.}\label{Fig:result_1}
\end{figure}
In the table~\ref{Tab.Method(I)}, we list $N=1,2, 5$, and $10$ results up to 15-loop including the tree ones, where the optimized values are shaded. Also we plot the ratio of approximate to exact values for $0 \leq L \leq 14$ (omitting the tree part) in the figure~\ref{Fig:result_1}. From these we convince the validity of the loop expansion in this case; since even in the smallest $N=1$ case, the 2- or 3-loop approximation gives $\sim 0.1$ or $\sim 0.2\%$ error. It also should be noted that the characteristic feature of the asymptotic expansion can be read from deviation after passing through the optimized values, which is most clearly seen in the figure~\ref{Fig:result_1} at the tail end of the $N=1$ line.


\noindent $\bullet $Method(II): from (\ref{Method2f}), the saddle point is
\begin{eqnarray}
 \tilde{f}'(t) = 1 - \left( 1-\frac{1}{N} \right) \frac{1}{t} =0   \Longrightarrow  t_c =  t_0 + \frac{t_1}{N} \ ; \quad t_0 \equiv 1 \ , \ t_1 \equiv  -1  \  , \label{t_c by t_0 t_1}
\end{eqnarray}
and satisfies the stability condition,
\begin{eqnarray}
\tilde{f}^{(2)}_c= \frac{1}{t_c} = \frac{N}{N-1} > 0   \ .   \label{StabCond.fTilde}
\end{eqnarray}
Note that
\begin{eqnarray}
   \tilde{f}_c = t_c\left( 1- \ln t_c \right)   \ ,  \quad  
  \tilde{f}^{(n)}_c \equiv \tilde{f}^{(n)}\! \left( t_c \right)= (-)^{n} (n-1)! \frac{1}{\left( t_c \right) ^{n-1}} \ ; \ \left( n \geq 2 \right) \ ,   \label{Tildef_0}
\end{eqnarray}
to give $\displaystyle{\tilde{F}_{n}= \frac{(-)^n (n-1)!}{t_c^{n/2-1}}}$ so that
\begin{eqnarray}
 \tilde{ F}(n_1, n_2, \dots , n_k) = (-)^{3k + \sum_j n_j }\frac{1}{ t_c^{ \left( \sum_jn_j + k \right) /2}} \prod_{j=1}^k  \frac{1}{(n_j+3)}= \frac{1}{\left(  t_c \right) ^L } \prod_{j=1}^k  \frac{1}{(n_j+3)}  \ , 
\end{eqnarray}
with $\sum_{j=1}^k n_j+ k =2 L$, from (\ref{TildeF Definition}). Therefore with the aid of (\ref{I_NTildeFinal})
\begin{eqnarray}
 I_N \approxeq {\mathrm e}^{-N t_c \left( 1 - \ln  t_c \right) }  \sqrt{\frac{2 \pi t_c}{N }}   \sum_{L=0}^\infty \frac{1}{N^L} \left(  t_c \right)^{- L }   \sum_{k=0}^{2L}  T \! \left( L,k |  2L-k\right)  \ . \label{I_N Method II}
\end{eqnarray}
The tree and the $1$-loop part are given
\begin{eqnarray}
& & \left(  I_N \right)_{\rm tree} = {\mathrm e}^{-N t_c \left( 1 - \ln  t_c \right) }\Big|_{t_1 =0} = {\mathrm e}^{-N}  \ , \\
& & \left(  I_N \right)_{1-\mbox{loop}} = {\mathrm e}^{-N t_c \left( 1 - \ln  t_c \right) }\sqrt{\frac{2  \pi t_c}{N }}  \Big|_{t_1 =0} = {\mathrm e}^{-N} \sqrt{\frac{2 \pi}{N }}  \ . 
\end{eqnarray}
From the $2$-loop approximation, $t_1$(\ref{t_c by t_0 t_1}) should be included to $t_c$, and all function of $t_c$ must be expanded up to $O(1/N^L): L=1,2 \dots $: in (\ref{I_N Method II}), introduce the prefactor ${\cal P}$,
\begin{eqnarray}
{\cal P} \equiv {\mathrm e}^{-N t_c \left( 1 - \ln  t_c \right) }  \sqrt{\frac{2 \pi t_c}{N }}= {\mathrm e}^{-N} \sqrt{\frac{2 \pi}{N }} \left[ 1 -\frac{1}{12} \frac{1}{N^2} -\frac{1}{12} \frac{1}{N^3}  -\frac{103}{1440} \frac{1}{N^4} + O \! \left( \frac{1}{N^5} \right) \right]  \ , \label{GammaPreFactor}
\end{eqnarray}
and the loop factor ${\cal L}$
\begin{eqnarray}
& & \hspace{-4ex}{\cal L} \equiv  \sum_{L=0}^\infty \frac{1}{N^L} \left(  t_c \right)^{- L }   \sum_{k=0}^{2L}  T \! \left( L,k |  2L-k\right) = 1 + \frac{n(1)}{N} + \frac{n(1) + n(2)}{N^2} \nonumber \\
& & \hspace{4ex} + \frac{n(1) + 2n(2) + n(3)}{N^3} + \frac{n(1) + 3n(2) + 3n(3) + n(4)}{N^4} +  O \! \left( \frac{1}{N^5} \right)  \ , 
\end{eqnarray}
where
\begin{eqnarray}
& & \hspace{8ex} n(L) \equiv  \sum_{k=0}^{2L}  T \! \left( L,k |  2L-k\right)   \ ; \\
& &   n(1) = \frac{1}{12} \ ,  \  n(2) = \frac{1}{288} \ ,  \  n(3) = - \frac{139}{51840} \ , \ n(4) = -\frac{571}{2488320 } \ ,
\end{eqnarray}
from the table~\ref{table:=2L-k:1} in the appendix \ref{sec:appendixA}. ${\cal P} \times {\cal L}$ gives the $5$-loop approximation of $I_N$
\begin{eqnarray}
& & \left(  I_N \right)_{5-\mbox{loop}}= {\mathrm e}^{-N} \sqrt{\frac{2 \pi}{N }}\left[ 1 +  \sum_{L=1}^4 \frac{n(L)}{N^L}   \right]  \ , 
\end{eqnarray}
which is exactly the same to the one ($L \mapsto 4$ in (\ref{GammaWithEquality})) in Method(I). Many terms in the numerator, $n(1),\dots, n(L-1)$ are canceled, leaving us only $n(L)$. These miracle cancellations occur for all orders of $1/N$, yielding the result that {\bf there is no difference between Method(I) and (II)} in this case.

The reason is rather easily figured out: because of the formula (\ref{FormulaGammaCase}) in the appendix~\ref{sec:appendixB}, we can put $g(t)\equiv 1/t$ in the integral $I_N$ (\ref{IntegralI_N}) to the unity, $g(t) \mapsto 1$. In other words, {\bf Method(I) is equivalent to Method(II)} in the Gamma function case.

\section{A Fermionic Case}\label{sec:Fermion}
Our target is a fermionic version of (\ref{Model1});
\begin{eqnarray}
Z  \equiv \int d^{N+1}\hat{\mbit{\xi}}d^{N+1}\hat{\mbit{\xi}}^* \exp \left[ -\hat{\mbit{\xi}}^*\cdot\mbit{\omega} \cdot \hat{\mbit{\xi}} + \frac{\lambda^2}{2 N} \left( \hat{\mbit{\xi}}^* \cdot \hat{\mbit{\xi}}\right)^2 \right]
\ , \label{FermionZ}
\end{eqnarray}
where $\hat{\mbit{\xi}}, \hat{\mbit{\xi}}^*(\mbit{\xi}, \mbit{\xi}^*)$ are $N+1(N)$-dimensional Grassmann variables,
\begin{eqnarray}
 \hat{\mbit{\xi}} \equiv \left( \xi_0, \mbit{\xi} \right) \  ,  \  \hat{\mbit{\xi}}^* \equiv \left( \xi_0^*, \mbit{\xi}^* \right) \ ;  \quad   \mbit{\xi} \equiv \left( \xi_1,\cdots, \xi_N \right)  \ , \ \mbit{\xi}^* \equiv \left( \xi_1^*, \cdots, \xi_N^*  \right) \ , 
\end{eqnarray}
with
\begin{eqnarray}
d^{N+1} \hat{\mbit{\xi}}\equiv d^{N} \mbit{\xi}d \xi_0  \equiv d \xi_N d \xi_{N-1} \cdots d \xi_0  \ , \quad   d^{N+1}\hat{\mbit{\xi}}^* \equiv d\xi_0^*  d^{N} \mbit{\xi}^* \equiv d\xi_0^* d\xi_1^*  \cdots d\xi_N^*     \ ,
\end{eqnarray}
and $(N+1) \times (N+1)$ matrix, $\mbit{\omega}$, has been given in (\ref{omegaMatrix}).
$Z$ is calculable by means of a standard Grassmann integration;
\begin{eqnarray}
\int d^n \mbit{\xi}d^n \mbit{\xi}^* \left( \mbit{\xi}^* \cdot \mbit{\xi}\right)^m  = \left( - \right)^n n! \delta_{mn} \ ,
\end{eqnarray}
to obtain 
\begin{eqnarray}
Z = \sum_{r=0}^{ \left[ \frac{N}{2}  \right] }  \frac{N!}{r!(N-2r)!}\left( \omega_0\omega +\frac{\lambda^2}{N}(N-2r) \right)(\omega)^{N-2r-1}\left( \frac{\lambda^2}{2 N} \right)^r    \ .  \label{FermionExactZ}
\end{eqnarray}
In this analysis, we assume that all parameters in this model are real and positive\footnote{Although we can see an interesting phenomenon when $\lambda^2 <0$ :the caustics emerge\cite{rf:kashiSaka2}.}, $\omega > 0,  \omega_0 > 0 ,  \lambda > 0$, and take $N=2$ with
\begin{eqnarray}
   0 \leq  \lambda  \leq 10 \ ;  \quad \omega_0 = 10^{2}\omega \  , \ \omega \  ,  \ 10^{-2} \omega \ .\label{parametersRange}
\end{eqnarray}
(The case, $\omega_0 = 10^2 \omega $ is a toy model of u-, d-, and s-quarks.)

Introducing an auxiliary field, in terms of
\begin{eqnarray}
1=\int_{-\infty}^\infty \frac{dy}{\sqrt{2\pi}}\exp \left[ -\frac{1}{2}\left( y+\frac{\lambda}{\sqrt{N}}\left( \hat{\mbit{\xi}}^*\cdot \hat{\mbit{\xi}}\right) \right)^2 \right] \  , 
\end{eqnarray}
into the target (\ref{FermionZ}), we obtain
\begin{eqnarray}
Z=\sqrt{\frac{N}{2\pi}}\int_{-\infty}^\infty dy\left( \omega_0+\lambda y \right) \exp \left[ -N \left( \frac{y^2}{2} - \ln \left( \omega +\lambda y \right)   \right) \right] \ ,
\end{eqnarray}
where we have performed the Grassmann Gaussian integration,
\begin{eqnarray}
\int d \xi d \xi^* {\mathrm e}^{- \omega \xi^* \xi }= \omega  \ , 
\end{eqnarray}
and $y$ has been scaled, $y \mapsto \sqrt{N}y$, as before. Now write 
\begin{eqnarray}
& &    Z = \sqrt{\frac{N}{2\pi}} I_N  \ ;  \qquad I_N \equiv \int dt \  g(t) \ {\mathrm e}^{ -N f(t)}   \ ,\label{Fermion I_N}   \\ 
&& g(t) \equiv \omega_0+\lambda t   \ ;  \qquad    f(t)  \equiv  \frac{t^2}{2} - \ln \left( \omega +\lambda t \right)   \ ,\label{f,gDef}
\end{eqnarray}
for Method(I) and
\begin{eqnarray}
& & \hspace{8ex}      I_N = \int dt \ {\mathrm e}^{ -N \tilde{f}(t)}  \ ,  \nonumber \\
& & \hspace{-4ex}\tilde{f}(t) \equiv f(t) - \frac{1}{N} \ln g(t) = \frac{t^2}{2} - \ln \left( \omega +\lambda t \right)  -\frac{1}{N} \ln \left( \omega_0 +\lambda t \right)   \ ,\label{I_N Method2}
\end{eqnarray}
for Method(II). (Here $y$ has been switched to $t$.)

\noindent $\bullet $Method(I): the saddle points are determined by
\begin{eqnarray}
\left. f^{(1)}(t) \right|_{t=t_0} = t_0-\frac{\lambda}{\omega +\lambda t_0}=0 \ . \label{FermionSaddles}
\end{eqnarray}
Here and hereafter the equation is called as the gap equation\cite{rf:NJLK}. If we introduce
\begin{eqnarray}
 \Omega_0  \equiv \omega +\lambda t_0 \ ,
\end{eqnarray}
(\ref{FermionSaddles}) becomes 
\begin{eqnarray}
\left( \Omega_0 \right)^2-\omega \Omega_0 -\lambda^2=0  \ , \label{SaddlePoint Method1}
\end{eqnarray}
yielding to two saddle points
\begin{eqnarray}
\Omega_0^{(\pm)} \equiv \frac{ \omega \pm \sqrt{\omega^2+ 4 \lambda^2} }{2} \ . \label{TwoSaddlePoints} 
\end{eqnarray}
The stability condition (\ref{stabilityCond}) in this case reads
\begin{eqnarray}
f^{(2)}_0 =1+\left( \frac{\lambda}{\Omega_0} \right)^2 =\frac{\left( \Omega_0 \right)^2+\lambda^2}{\left( \Omega_0 \right)^2} > 0  \  ; 
\end{eqnarray}
which is positive for both $\Omega_0^{(\pm)}$. The value of $f(t), g(t)$ and derivatives at the saddle points are given by
\begin{eqnarray}
f_0 =\frac{\left( \Omega_0 -\omega  \right)^2}{2 \lambda^2}-\ln \Omega_0  &  ;  & f_0^{(m)}=(m-1)! \left( -\frac{\lambda}{\Omega_0} \right)^m \ ;\  (m \geq 3) \ ,\\ 
g_0 = \Omega_0 + \delta \omega & ; & g^{(1)}_0=\lambda \ ; \qquad   g^{(m)}_0 = 0 \ ;  \ (m \geq 2) \ ,
\end{eqnarray}
where
\begin{eqnarray}
\delta \omega \equiv  \omega_0-\omega \  . \label{delta omega Def}
\end{eqnarray}
Then from (\ref{F_nDefinition})
\begin{eqnarray}
\frac{F_{n_j+3}}{(n_j+3)!}= \frac{1}{(n_j+3)!} \left( f^{(2)}_0 \right)^{-(n_j+3)/2} f^{(n_j+3)}_0 
= \frac{1}{n_j+3} \left( \frac{ - \epsilon \! \left( \Omega_0  \right)  \lambda}{\sqrt{\left( \Omega_0 \right)^2+\lambda^2}} \right)^{n_j+3} \ ,
\end{eqnarray}
where 
\begin{eqnarray}
 \epsilon \! \left( \Omega_0  \right)  = \left\{
                                       \begin{array}{rc}
                                        1  \ : 	&   \Omega_0  > 0    \\ 
                                        -1 \ : &    \Omega_0  < 0
                                       \end{array}
                                       \right.     , 
\end{eqnarray}
is the sign function.
Accordingly, in view of (\ref{I_NFinal}) with (\ref{F Definition}), we find
\begin{eqnarray}
& &  \hspace{-6ex}\frac{(-)^k}{k!} \left( 2(L+k) -1\right)!! \  \sum_{ \mbox{ all possible} \ \left\{ n_j \right\} }^{ \sum_{j=1}^k n_j \leq 2L-k  }   \hspace{-2ex}   F(n_1, n_2, \dots , n_k)     \nonumber \\  
& & \hspace{-4ex} =   \left( \frac{\lambda }{\sqrt{\left( \Omega_0 \right)^2 + \lambda^2 }} \right)^{2(L+k)}  \hspace{-1ex}  \left[  \left( \Omega_0 + \delta \omega   \right)  T \! \left( L,k | 2L-k \right) - \Omega_0   T \! \left( L,k | 2L-k-1 \right)  \right]  \ ,
\end{eqnarray}
where use has been made of the notation (\ref{T(L,k| )Def.n_k}): $ T \! \left( L,k | 2L-k-1 \right) $ is defined by replacing the sum $\sum_{j=1}^k n_j = 2L-k$ to $\sum_{j=1}^k n_j = 2L-k-1$ in (\ref{T(L,k| )Def.n_k}). Therefore from (\ref{F Definition})
\begin{eqnarray}
& &  \hspace{-4ex}   I_N  \approxeq  \sqrt{\frac{2 \pi }{N}}\epsilon (\Omega_0)   \frac{\Omega_0^{N+1}}{\sqrt{\left( \Omega_0 \right)^2+\lambda^2}}     \exp \left[ - N \frac{\left( \Omega_0 -\omega  \right)^2}{2 \lambda^2} \right]     \sum_{L=0}^\infty  \frac{1}{N^L} \sum_{k=0}^{2L}   \left( \frac{\lambda }{\sqrt{\left( \Omega_0 \right)^2 + \lambda^2 }} \right)^{2(L+k)}  \nonumber \\ 
& &  \hspace{8ex}   \times     \left[  \left( \Omega_0 + \delta \omega   \right)  T \! \left( L,k | 2L-k \right) - \Omega_0   T \! \left( L,k | 2L-k-1 \right)  \right] \ ,  
\end{eqnarray} 
so that 
\begin{eqnarray}
& &  \hspace{0mm}   Z  \approxeq  \epsilon (\Omega_0)   \frac{\Omega_0^{N+1}}{\sqrt{\left( \Omega_0 \right)^2+\lambda^2}}     \exp \left[ - N \frac{ \left( \Omega_0 -\omega  \right)^2}{2 \lambda^2}  \right]     \sum_{L=0}^\infty  \frac{1}{N^L} \sum_{k=0}^{2L}   \left( \frac{\lambda }{\sqrt{\left( \Omega_0 \right)^2 + \lambda^2 }} \right)^{2(L+k)}  \nonumber \\ 
& &  \hspace{8ex}   \times     \left[  \left( \Omega_0 + \delta \omega   \right)  T \! \left( L,k | 2L-k \right) - \Omega_0   T \! \left( L,k | 2L-k-1 \right)  \right] \ .   
\end{eqnarray} 
According to classification in sec.\ref{sec.SaddlePoints}, (\ref{TreePart Def}) and (\ref{L-loop Def}), the tree and the $l$-loop approximation read
\begin{eqnarray}
& & Z_{\rm tree} \equiv {\mathrm e}^{-N f_0} g_0 =  \exp \left[ - N \frac{\left( \Omega_0 -\omega  \right)^2}{2 \lambda^2}   \right]\Omega_0^N \left( \Omega_0 + \delta \omega  \right)  \ ,  \label{FermiTreeMethodI}  \\
& & \hspace{-3ex}Z_{ l \mbox{-loop}} \equiv  \epsilon (\Omega_0)   \frac{\Omega_0^{N+1}}{\sqrt{\left( \Omega_0 \right)^2+\lambda^2}}     \exp \left[ -N \frac{\left( \Omega_0 -\omega  \right)^2}{2 \lambda^2} \right]     \sum_{L=0}^{l-1}  \frac{1}{N^L} \sum_{k=0}^{2L}   \left( \frac{\lambda }{\sqrt{\left( \Omega_0 \right)^2 + \lambda^2 }} \right)^{2(L+k)}  \nonumber \\ 
& &  \hspace{8ex}   \times     \left[  \left( \Omega_0 + \delta \omega   \right)  T \! \left( L,k | 2L-k \right) - \Omega_0   T \! \left( L,k | 2L-k-1 \right)  \right]  \ . \label{Z l-loop}
\end{eqnarray}
Up to $3$-loop ($L \leq 2$), by noting the table~\ref{table:=2L-k:1} in the appendix~\ref{sec:appendixA} and $T \! \left( 1, 1 | 0 \right) =-1, T \! \left( 2, 1 | 2 \right) =-3, T \! \left( 2, 2 | 1 \right) =35/4,  T \! \left( 2, 3 | 1 \right) =-35/6$ for $ T \! \left( L,k | 2L-k-1 \right) $, we find
\begin{eqnarray}
& & Z_{ 1 \mbox{-loop}}  =   \epsilon (\Omega_0)  \exp \left[ - N \frac{\left( \Omega_0 -\omega  \right)^2}{2 \lambda^2}   \right] \frac{\left( \Omega_0 \right)^{N+1} \left( \Omega_0 + \delta \omega  \right)}{\sqrt{\left( \Omega_0 \right)^2+ \lambda^2  }} \ , \label{Fermi1-loopMethodI} \\ 
& & Z_{ 2 \mbox{-loop}}  = Z_{ 1 \mbox{-loop}}+   \epsilon (\Omega_0)   \exp \left[ - N \frac{\left( \Omega_0 -\omega  \right)^2}{2 \lambda^2}   \right]  \frac{\left( \Omega_0 \right)^{N+1} }{\sqrt{\left( \Omega_0 \right)^2+ \lambda^2  }} \nonumber \\ 
& & \hspace{8ex}   \times \frac{1}{N}\left[ \frac{\lambda^4 \left( \Omega_0 -3 \delta \omega  \right) }{4\left[ \left( \Omega_0 \right)^2+ \lambda^2   \right]^2  }  +  \frac{5 \lambda^6  \left( \Omega_0 + \delta \omega  \right) }{6\left[ \left( \Omega_0 \right)^2+ \lambda^2   \right]^3  }\right]  \ , \label{Fermi2-loopMethodI}
\end{eqnarray}
and
\begin{eqnarray} 
& &\hspace{-2ex} Z_{ 3 \mbox{-loop}}  = Z_{ 2 \mbox{-loop}} +   \epsilon (\Omega_0)   \exp \left[ - N \frac{\left( \Omega_0 -\omega  \right)^2}{2 \lambda^2}   \right]  \frac{\left( \Omega_0 \right)^{N+1} }{\sqrt{\left( \Omega_0 \right)^2+ \lambda^2  }}\frac{1}{N^2} \left[ \frac{ \lambda^6  \left( \Omega_0 -5 \delta \omega  \right) }{2 \left[ \left( \Omega_0 \right)^2+ \lambda^2   \right]^3  } \right.  \nonumber \\ 
& & \hspace{2ex} \left.    +   \frac{ 7 \lambda^8 \left( 7\Omega_0 +47 \delta \omega  \right) }{32\left[ \left( \Omega_0 \right)^2+ \lambda^2   \right]^4  } -   \frac{3 5 \lambda^{10} \left( 5 \Omega_0 + 9 \delta \omega  \right) }{24 \left[ \left( \Omega_0 \right)^2+ \lambda^2   \right]^5   } + \frac{38 5 \lambda^{12} \left(  \Omega_0 + \delta \omega  \right) }{72 \left[ \left( \Omega_0 \right)^2+ \lambda^2   \right]^6   }     \right]  \ .  
\end{eqnarray}
Since there are two saddle points $\Omega_0^{\pm }$ (\ref{TwoSaddlePoints}) the total $Z$ is given by
\begin{eqnarray}
Z_{l \mbox{-loop}}^{\rm Total} = Z_{ l \mbox{-loop}}^{(+)} + Z_{ l \mbox{-loop}}^{(-)}  \ ,
\end{eqnarray}
where $Z_{ l \mbox{-loop}}^{(\pm) }$ has been obtained by putting $\Omega_0 \mapsto \Omega_0^{(\pm) }$ in (\ref{Z l-loop}).

In the table \ref{table: FermiMethod(I)}, we list the result of $\omega_0 = 10^{2} \omega , \omega , 10^{-2}\omega$ for $10^{-3} \leq \lambda \leq 10$ in $N=2$. We put $\omega \mapsto 1$ and write the data of the ratio of $Z_{\rm tree}$ and $Z_{l-\mbox{loop}}(l=1,2,3)$ to the exact value. From this, in the weak coupling region, $\lambda < 1$, the $1$-loop approximation almost yields the exact value; even in the worst case, $\omega_0 = 10^{-2}$, only $0.3\%$ error crops up. For a whole coupling region including $\lambda \geq 1$, the error is within $1.1\%$ under $2$-loop and becomes venial, $< 0.3 \% $, under the $3$-loop approximation.
\begin{table}[h]
{\tiny 
\begin{eqnarray*}
\begin{array}{|l|cccccc|}
\hline\hline
 \omega_0 = 10^{2}
 &\lambda & \textrm{Exact} &
 \begin{array}{c} \textrm{ tree}\\ \textrm{ (tree)/Exact}\\ \end{array} &
 \begin{array}{c} \textrm{ 1-loop}\\ \textrm{ (1-loop)/Exact}\\ \end{array} &
 \begin{array}{c} \textrm{ 2-loop}\\ \textrm{ (2-loop)/Exact}\\ \end{array} &
 \begin{array}{c} \textrm{ 3-loop}\\ \textrm{ (3-loop)/Exact}\\ \end{array} \\
\cline{2-7}
 &10^{-3} & 1.0000\times 10^{2} &
  \begin{array}{c} 1.0000\times 10^{2} \\ 1.0000 \\ \end{array} &
  \begin{array}{c} 1.0000\times 10^{2} \\ 1.0000 \\ \end{array} &
  \begin{array}{c} 1.0000\times 10^{2} \\ 1.0000 \\ \end{array} &
  \begin{array}{c} 1.0000\times 10^{2} \\ 1.0000 \\ \end{array} \\
\gcline{2-7}
 &10^{-2} & 1.0001\times 10^{2} &
  \begin{array}{c} 1.0001\times 10^{2} \\ 1.0000 \\ \end{array} &
  \begin{array}{c} 1.0001\times 10^{2} \\ 1.0000 \\ \end{array} &
  \begin{array}{c} 1.0001\times 10^{2} \\ 1.0000 \\ \end{array} &
  \begin{array}{c} 1.0001\times 10^{2} \\ 1.0000 \\ \end{array} \\
\gcline{2-7}
 &10^{-1} & 1.0051\times 10^{2} &
  \begin{array}{c} 1.0101\times 10^{2} \\ 1.0049 \\ \end{array} &
  \begin{array}{c} 1.0051\times 10^{2} \\ 1.0000 \\ \end{array} &
  \begin{array}{c} 1.0051\times 10^{2} \\ 1.0000 \\ \end{array} &
  \begin{array}{c} 1.0051\times 10^{2} \\ 1.0000 \\ \end{array} \\
\gcline{2-7}
 &1       & 1.5100\times 10^{2} &
  \begin{array}{c} 1.8253\times 10^{2} \\ 1.2088 \\ \end{array} &
  \begin{array}{c} 1.5438\times 10^{2} \\ 1.0224 \\ \end{array} &
  \begin{array}{c} 1.5138\times 10^{2} \\ 1.0025 \\ \end{array} &
  \begin{array}{c} 1.5096\times 10^{2} \\ 0.9998 \\ \end{array} \\
\gcline{2-7}
 &10      & 5.2000\times 10^{3} &
  \begin{array}{c} 7.5782\times 10^{3} \\ 1.4574 \\ \end{array} &
  \begin{array}{c} 5.3961\times 10^{3} \\ 1.0377 \\ \end{array} &
  \begin{array}{c} 5.1861\times 10^{3} \\ 0.9973 \\ \end{array} &
  \begin{array}{c} 5.1906\times 10^{3} \\ 0.9982 \\ \end{array} \\
\hline\hline
 \omega_0 =  \omega
 &\lambda & \textrm{Exact} &
 \begin{array}{c} \textrm{ tree}\\ \textrm{ (tree)/Exact}\\ \end{array} &
 \begin{array}{c} \textrm{ 1-loop}\\ \textrm{ (1-loop)/Exact}\\ \end{array} &
 \begin{array}{c} \textrm{ 2-loop}\\ \textrm{ (2-loop)/Exact}\\ \end{array} &
 \begin{array}{c} \textrm{ 3-loop}\\ \textrm{ (3-loop)/Exact}\\ \end{array} \\
\cline{2-7}
 &10^{-3} & 1.0000 &
  \begin{array}{c} 1.0000 \\ 1.0000 \\ \end{array} &
  \begin{array}{c} 1.0000 \\ 1.0000 \\ \end{array} &
  \begin{array}{c} 1.0000 \\ 1.0000 \\ \end{array} &
  \begin{array}{c} 1.0000 \\ 1.0000 \\ \end{array} \\
\gcline{2-7}
 &10^{-2} & 1.0002 &
  \begin{array}{c} 1.0002 \\ 1.0000 \\ \end{array} &
  \begin{array}{c} 1.0001 \\ 1.0000 \\ \end{array} &
  \begin{array}{c} 1.0001 \\ 1.0000 \\ \end{array} &
  \begin{array}{c} 1.0002 \\ 1.0000 \\ \end{array} \\
\gcline{2-7}
 &10^{-1} & 1.0150 &
  \begin{array}{c} 1.0200 \\ 1.0049 \\ \end{array} &
  \begin{array}{c} 1.0150 \\ 1.0000 \\ \end{array} &
  \begin{array}{c} 1.0150 \\ 1.0000 \\ \end{array} &
  \begin{array}{c} 1.0150 \\ 1.0000 \\ \end{array} \\
\gcline{2-7}
 &1       & 2.5000 &
  \begin{array}{c} 2.8740 \\ 1.1496 \\ \end{array} &
  \begin{array}{c} 2.4503 \\ 0.9801 \\ \end{array} &
  \begin{array}{c} 2.4934 \\ 0.9974 \\ \end{array} &
  \begin{array}{c} 2.4998 \\ 0.9999 \\ \end{array} \\
\gcline{2-7}
 &10      & 1.5100\times 10^{2} &
  \begin{array}{c} 1.8498\times 10^{2} \\ 1.2250 \\ \end{array} &
  \begin{array}{c} 1.4410\times 10^{2} \\ 0.9543 \\ \end{array} &
  \begin{array}{c} 1.5030\times 10^{2} \\ 0.9954 \\ \end{array} &
  \begin{array}{c} 1.5124\times 10^{2} \\ 1.0016 \\ \end{array} \\
\hline\hline
\omega_0 = 10^{-2}
 &\lambda & \textrm{Exact} &
 \begin{array}{c} \textrm{ tree}\\ \textrm{ (tree)/Exact}\\ \end{array} &
 \begin{array}{c} \textrm{ 1-loop}\\ \textrm{ (1-loop)/Exact}\\ \end{array} &
 \begin{array}{c} \textrm{ 2-loop}\\ \textrm{ (2-loop)/Exact}\\ \end{array} &
 \begin{array}{c} \textrm{ 3-loop}\\ \textrm{ (3-loop)/Exact}\\ \end{array} \\
\cline{2-7}
 &10^{-3} & 1.0001\times 10^{-2} &
  \begin{array}{c} 1.0001\times 10^{-2} \\ 1.0000 \\ \end{array} &
  \begin{array}{c} 1.0001\times 10^{-2} \\ 1.0000 \\ \end{array} &
  \begin{array}{c} 1.0001\times 10^{-2} \\ 1.0000 \\ \end{array} &
  \begin{array}{c} 1.0001\times 10^{-2} \\ 1.0000 \\ \end{array} \\
\gcline{2-7}
 &10^{-2} & 1.0101\times 10^{-2} &
  \begin{array}{c} 1.0101\times 10^{-2} \\ 1.0000 \\ \end{array} &
  \begin{array}{c} 1.0100\times 10^{-2} \\ 1.0000 \\ \end{array} &
  \begin{array}{c} 1.0100\times 10^{-2} \\ 1.0000 \\ \end{array} &
  \begin{array}{c} 1.0100\times 10^{-2} \\ 1.0000 \\ \end{array} \\
\gcline{2-7}
 &10^{-1} & 2.0050\times 10^{-2} &
  \begin{array}{c} 2.0100\times 10^{-2} \\ 1.0025 \\ \end{array} &
  \begin{array}{c} 2.0002\times 10^{-2} \\ 0.9976 \\ \end{array} &
  \begin{array}{c} 2.0049\times 10^{-2} \\ 1.0000 \\ \end{array} &
  \begin{array}{c} 2.0050\times 10^{-2} \\ 1.0000 \\ \end{array} \\
\gcline{2-7}
 &1       & 1.0150 &
  \begin{array}{c} 1.0774 \\ 1.0615 \\ \end{array} &
  \begin{array}{c} 0.9311 \\ 0.9173 \\ \end{array} &
  \begin{array}{c} 1.0046 \\ 0.9897 \\ \end{array} &
  \begin{array}{c}  1.0152 \\  1.0002 \\ \end{array} \\
\gcline{2-7}
 &10      & 1.0051\times 10^{2} &
  \begin{array}{c} 1.1105\times 10^{2} \\ 1.1049 \\ \end{array} &
  \begin{array}{c} 9.1577\times 10^{1} \\ 0.9111 \\ \end{array} &
  \begin{array}{c} 9.9945\times 10^{1} \\ 0.9944 \\ \end{array} &
  \begin{array}{c} 1.0084\times 10^{2} \\ 1.0033 \\ \end{array} \\
\hline\hline
\end{array}
\end{eqnarray*}

\vspace{-4ex}

\caption{Result of Method(I): $N=2$, $\omega=1$ and $\omega_0 = 10^{2}, 1, 10^{-2}$ for $10^{-3} \leq \lambda \leq 10$. The error is within $1.1(0.3)\%$ under the 2(3)-loop approximation for a whole coupling region.}\label{table: FermiMethod(I)}
}  
\end{table}

\noindent $\bullet $Method(II): in (\ref{I_N Method2}), the saddle points are given by the gap equation,
\begin{eqnarray}
\left. \tilde{f}^{(1)}(t) \right|_{t_c} = t_c - \frac{\lambda }{\omega + \lambda t_c} - \frac{1}{N}\frac{\lambda }{\omega_0 + \lambda t_c} = 0 \ ,
\end{eqnarray}
yielding to
\begin{eqnarray}
 \Omega_c - \omega - \frac{\lambda^2 }{\Omega_c} - \frac{1}{N}\frac{\lambda^2 }{\Omega_c + \delta \omega } = 0 \ ,   \qquad 
\Omega_c \equiv \omega + \lambda t_c  \ , \label{SaddlePoint Method2}
\end{eqnarray}
with $\delta \omega $ being given by (\ref{delta omega Def}). This is a cubic equation of $\Omega_c$, 
\begin{eqnarray}
\left( \Omega_c + \delta \omega  \right) \left( \left( \Omega_c \right)^2 - \omega  \Omega_c  - \lambda^2 \right)  = \frac{\lambda^2 }{N} \Omega_c  \ ,\label{GapEq Method2}
\end{eqnarray}
contrary to Method(I), where it was quadratic, (\ref{SaddlePoint Method1}). We write
\begin{eqnarray}
\tilde{f}^{(n)}(t_c) \equiv \tilde{f}^{(n)}_c    \ . 
\end{eqnarray}
The stability condition is fulfilled, 
\begin{eqnarray}
\tilde{f}^{(2)}_c = 1 + \frac{\lambda^2 }{\left( \Omega_c \right)^2} + \frac{1}{N}\frac{\lambda^2 }{\left( \Omega_c + \delta \omega  \right)^2}  > 0  \  , 
\end{eqnarray}
for any (three) saddle points $\Omega_c$. The values of $\tilde{f}(t)$ and derivatives at $\Omega_c$ are given
\begin{eqnarray}
& & \hspace{5ex} \tilde{f}_c= \frac{1}{2\lambda^2}\left( \Omega_c-\omega \right)^2-\ln \Omega_c - \frac{1}{N} \ln (\Omega_c+\delta \omega) \ , \\ 
& &  \tilde{f}^{(n)}_c = (-)^n (n-1)!  \left[  \left(\frac{\lambda}{\Omega_c} \right)^n + \frac{1}{N} \left( \frac{\lambda}{ \Omega_c + \delta \omega  }\right)^n    \right] \ ; \quad  n \geq 3  \ .  
\end{eqnarray} 
Then from (\ref{I_NTildeFinal}) 
\begin{eqnarray}
& & \hspace{-5ex} \frac{\tilde{F}_{n_j+3}}{(n_j+3)!}  \nonumber \\ 
 & & \hspace{-2ex}=\frac{\left[ -  \lambda \epsilon(\Omega_c)\epsilon(\Omega_c + \delta \omega )  \right]^{n_j+3} }{n_j+3}   \frac{  \left( \Omega_c+ \delta \omega  \right)^{n_j + 3} + \frac{1}{N}\left( \Omega_c \right)^{n_j+3}  }{ \left[ \left( \left( \Omega_c  \right)^2 + \lambda^2  \right) \left( \Omega_c+ \delta \omega \right)^2 +  \frac{\lambda^2 }{N} \left( \Omega_c  \right)^{2} \right]^{(n_j+3)/2} }  \ , 
\end{eqnarray}
to give, with using the condition $\sum_{j=1}^k n_j = 2L-k$, 
\begin{eqnarray}
& &\hspace{-3ex} I_N  \approxeq  \sqrt{\frac{2 \pi }{N}}   \exp \left[ - \frac{ N \left( \Omega_c - \omega  \right)^2}{2 \lambda^2 } \right] \frac{  \left( \Omega_c  \right)^N \left( \Omega_c + \delta \omega  \right) }{\sqrt{1 + \left( \lambda / \Omega_c  \right)^2 + \frac{1}{N} \left( \lambda / ( \Omega_c + \delta \omega )  \right) ^2   }} \nonumber \\
& & \hspace{0ex}\times \sum_{L=0}^\infty  \frac{1}{N^L} \sum_{k=0}^{2L} \frac{(-)^k }{ k! }\left( 2(L+k)-1 \right)!!  \frac{\lambda^{2(L+k)}}{ \left[  \left( \left( \Omega_c  \right)^2 + \lambda^2  \right) \left( \Omega_c+ \delta \omega \right)^2 +  \frac{\lambda^2 }{N} \left( \Omega_c  \right)^{2}   \right]^{L+k}  }      \nonumber \\ 
& & \hspace{5ex}  \times  \hspace{-2ex} \sum_{ \mbox{ all possible} \ \left\{ n_j \right\} }^{ \sum_{j=1}^k n_j = 2L-k  }   \prod_{j=1}^k  \frac{1}{n_j+3} \left[  \left( \Omega_c+ \delta \omega  \right)^{n_j + 3} + \frac{1}{N}\left( \Omega_c \right)^{n_j+3}  \right]   \  .   \label{I_N Fermion Method2}      
\end{eqnarray}
$Z$, (\ref{Fermion I_N}), is expressed, therefore, by a product of a prefactor ${\cal P}$ and a loop factor ${\cal L}$, 
\begin{eqnarray}
 Z  \equiv  {\cal P} \times {\cal L}  \ ,  \label{Z by P&L}
\end{eqnarray}
with
\begin{eqnarray}
& & {\cal P} \equiv  \exp \left[ - \frac{ N \left( \Omega_c - \omega  \right)^2}{2 \lambda^2 } \right] \frac{  \left( \Omega_c  \right)^N \left( \Omega_c + \delta \omega  \right) }{\sqrt{1 + \left( \lambda / \Omega_c  \right)^2 + \frac{1}{N} \left( \lambda / ( \Omega_c + \delta \omega )  \right) ^2   }}   \  ;  \label{PreFactor Def}   \\
& & {\cal L} \equiv \sum_{L=0}^\infty  \frac{1}{N^L} \sum_{k=0}^{2L} \frac{(-)^k }{ k! }\left( 2(L+k)-1 \right)!!  \frac{\lambda^{2(L+k)}}{ \left[  \left( \left( \Omega_c  \right)^2 + \lambda^2  \right) \left( \Omega_c+ \delta \omega \right)^2 +  \frac{\lambda^2 }{N} \left( \Omega_c  \right)^{2}   \right]^{L+k}  }      \nonumber \\ 
& & \hspace{5ex}  \times  \hspace{-2ex}  \sum_{ \mbox{ all possible} \ \left\{ n_j \right\} }^{ \sum_{j=1}^k n_j = 2L-k  }  \prod_{j=1}^k  \frac{1}{n_j+3} \left[  \left( \Omega_c+ \delta \omega  \right)^{n_j + 3} + \frac{1}{N}\left( \Omega_c \right)^{n_j+3}  \right]   \ . \label{LoopFactor Def}
\end{eqnarray}

Now solve the gap equation to find that there are three kinds of $1/N$ series, 
\begin{eqnarray}
\Omega_c^{(i)} =   \Omega_0^{(i)} + \sum_{j=1} \frac{\Omega_j^{(i)}}{N^j}  \ ; \quad (i=1,2,3)  \ .  \label{Omega_cExpansion} 
\end{eqnarray}
where
\begin{eqnarray}
 \Omega_0^{(i)} = \left\{
                                                \begin{array}{rc}
                                                \Omega_0^{(\pm ) }  \  ;  &  \hat{i}=1,2    	\\
                                                \noalign{\vspace{1ex}}
                                               -  \delta \omega   \  ;  &  i=3   
                                                \end{array}
                                                \right.  \ ,  \label{Omega_cLeading}
\end{eqnarray}
with $ \Omega_0^{(\pm ) }$ being given by (\ref{TwoSaddlePoints}), then up to $O \! \left( 1/N^2 \right) $,
\begin{eqnarray}
& &  \Omega_1^{(i)} = \frac{ \lambda^2  \Omega_0^{(i)} }{ B_i }  \ ;  \qquad  \Omega_2^{(i)} = \frac{\lambda^2  \Omega_1^{(i)} -  \left( \Omega_1^{(i)} \right)^2 A_i }{B_i} \ .  \label{Omega_1}    \\
& & A_i \equiv 3 \Omega_0^{(i)} - \omega + \delta \omega \ ; \quad  B_i \equiv 3 \left( \Omega_0^{(i)} \right)^2 - 2(\omega - \delta \omega)\Omega_0^{(i)} - \omega  \delta \omega - \lambda^2  \ .  \label{A&B}
\end{eqnarray}
These are sufficient under the 3-loop approximation ($O(1/N^2)$). Accordingly write $Z$ as $Z^{(i)}$ in (\ref{Z by P&L}) such that
\begin{eqnarray}
Z^{(i)}= {\cal P}^{(i)} \times {\cal L}^{(i)}  \  ;   \quad (i=1,2,3)  \ , \label{Z^i by P^i L^i}
\end{eqnarray}
with the prefactor, 
\begin{eqnarray}
 {\cal P}^{(i)} \equiv F^{(i)}(1) F^{(i)}(2) F^{(i)}(3) \left( \Omega^{(i)}_{c} + \delta \omega  \right)  \ ,\label{P by F(1),,,F(3)}  
 \end{eqnarray}
where
\begin{eqnarray}
 & & \hspace{-8ex}F^{(i)}(1) \equiv \exp \left[ - N \frac{\left( \Omega_c^{(i)} - \omega \right)^2 }{2 \lambda^2 } \right]   \ ; \qquad  F^{(i)}(2) \equiv \left( \Omega_c^{(i)} \right)^N  \ , \label{F(1)F(2) Def} \\
& &  F^{(i)}(3) \equiv \left[ 1 + \left( \frac{\lambda}{\Omega_c^{(i)} }   \right)^2 + \frac{1}{N} \left( \frac{\lambda}{\Omega_c^{(i)} + \delta \omega}  \right) ^2  \right]^{-1/2}  \ ,
\label{F(3) Def}
\end{eqnarray}
and the loop factor, ${\cal L}^{(i)} $, up to $O(1/N^2)$,
\begin{eqnarray}
& &\hspace{-4ex} {\cal L}^{(i)} \equiv \sum_{L=0}^\infty  \frac{1}{N^L} \sum_{k=0}^{2L} \frac{(-)^k }{ k! }\left( 2(L+k)-1 \right)!! \left( \frac{\lambda^2 }{ F^{(i)} (4) } \right)^{ \hspace{-1ex}L+k} \hspace{-1ex} \sum_{ \mbox{ all possible} \ \left\{ n_j \right\} }^{ \sum_{j=1}^k n_j = 2L-k  }   \prod_{j=1}^k  \frac{F^{(n_j +3  ; i)}(5) }{n_j+3}    \nonumber \\
& & \hspace{0ex}=  1   + \frac{\lambda^2}{N  F^{(i)}(4) }  \left[   - \frac{3}{4} \frac{ \lambda^2 F^{(4;i)}(5)}{ F^{(i)}(4)} + \frac{5}{6} \left(  \frac{ \lambda^2 F^{(3;i)}(5)   }{  F^{(i)}(4)  } \right)^2  \right]   \nonumber \\
& &   \hspace{4ex} + \frac{\lambda^4}{\left( N   F^{(i)}(4) \right)^2  } \left[   - \frac{5}{2}  \frac{ \lambda^2 F^{(6;i)}(5)}{ F^{(i)}(4)}  +  7 \frac{  \lambda^4 F^{(3;i)}(5) F^{(5;i)}(5) }{ \left(  F^{(i)}(4) \right)^2 }  +  \frac{105}{32}   \left(  \frac{\lambda^2 F^{(4;i)}(5)}{F^{(i)}(4)}   \right)^2  \right.  \nonumber \\
& & \hspace{8ex} \left.  -   \frac{105}{8} \frac{ \lambda^6 \left(  F^{(3;i)}(5)   \right)^2  F^{(4;i)}(5) }{\left(    F^{(i)}(4)   \right)^3}   +    \frac{385}{72} \left(    \frac{ \lambda^2 F^{(3;i)}(5)   }{  F^{(i)}(4) }\right)^4  \right] \ ,   \label{L by F(4) & F(5)} 
\end{eqnarray}
where
\begin{eqnarray}
& & F^{(i)}(4) \equiv \left( \left( \Omega_c^{(i)}  \right)^2 + \lambda^2  \right) \left( \Omega_c^{(i)} + \delta \omega \right)^2 +  \frac{\lambda^2 }{N} \left( \Omega_c^{(i)}  \right)^{2}   \  ;  \label{F(4) Def} \\ 
& &  F^{(M  ; i)}(5) \equiv  \left( \Omega_c^{(i)} + \delta \omega  \right)^{M} + \frac{1}{N}\left( \Omega_c^{(i)} \right)^{M} \ ; \ M=3, 4, \dots, 6 \ . \label{F(5) Def}
\end{eqnarray}
Note that 
\begin{eqnarray}
\Omega_c^{(i)}+ \delta \omega =\left\{
                               \begin{array}{cc}
 \displaystyle{\left( \Omega_0^{(\hat{i})}+ \delta \omega \right) \left[ 1 + \frac{1}{N} \frac{\Omega_1^{(\hat{i})}  }{\Omega_0^{(\hat{i})}+ \delta \omega} +  \frac{1}{N^2} \frac{\Omega_2^{(\hat{i})}  }{\Omega_0^{(\hat{i})}+ \delta \omega} \right] }\ ;  &  \hat{i}= 1,2  \\  
 \noalign{\vspace{1ex}}
 \displaystyle{ \hspace{5ex}\frac{\Omega_1^{(3)}}{N}\left[ 1 + \frac{1}{N}\frac{\Omega_2^{(3)}}{\Omega_1^{(3)}} \right]   }\ ;  &  i=3                                \end{array}
                               \right.  \ , 
\end{eqnarray}
so that the contribution of the third saddle, $i=3$, starts from $O \! \left( 1/N \right)$, that is, $2$-loop. Therefore we write
\begin{eqnarray}                                 
{\cal P}^{(i)} =\left\{
                    \begin{array}{cl}
       \displaystyle{{\cal P}_0^{(\hat{i})} \left[ 1 + \frac{ {\cal P}_1^{(\hat{i})}}{N} + \frac{{\cal P}_2^{(\hat{i})} }{N^2 } \right] }                	&  ; \  \hat{i} =1,2	\\
       \noalign{\vspace{1ex}}
     \displaystyle{  \frac{ {\cal P}_1^{(3)}}{N} \left[ 1  + \frac{{\cal P}_2^{(3)} }{N } \right]   }                   	& ; \  i=3
                    \end{array}
                    \right.   \ ; \label{P'sExpansion} 
\end{eqnarray}
and
\begin{eqnarray} 
 {\cal L}^{(i)} =\left\{
                 \begin{array}{cl}
       \displaystyle{{\cal L}_0^{(\hat{i})} + \frac{ {\cal L}_1^{(\hat{i})} }{N} + \frac{ {\cal L}_2^{(\hat{i})}}{N^2}}            	& ; \  \hat{i} =1,2	\\
    \displaystyle{{\cal L}_0^{(3)} + \frac{ {\cal L}_1^{(3)} }{N}  }              	&  ; \ i=3  
                 \end{array}
                 \right. \ .  \label{L'sExpansion}
\end{eqnarray}
Explicit forms of those functions, ${\cal P}_0^{(\hat{i})} \sim {\cal P}_1^{(3)}$ and ${\cal L}_0^{(\hat{i})} \sim {\cal L}_1^{(3)} $, are (after lengthy calculation) given in the appendix~\ref{sec:appendixC}.

The tree part is, from (\ref{F_0(1)}) and (\ref{F_0(2) F_1(2)}),
\begin{eqnarray}
& &  Z_{\rm  tree}^{(\hat{i})} =  F_0^{(\hat{i})}(1) F_0^{(\hat{i})}(2)\left( \Omega_0^{(\hat{i})} + \delta \omega  \right) \Bigg|_{\Omega_1^{(\hat{i})} \mapsto 0 } \nonumber \\
& & \hspace{5ex} =  \exp \left[ -\frac{N}{2 \lambda^2 } \left( \Omega_0^{(\hat{i})} - \omega  \right)^2  \right] \! \! \left(  \Omega_0^{(\hat{i})} \right)^N \! \! \left( \Omega_0^{(\hat{i})} + \delta \omega  \right) \  .
\end{eqnarray}
Since $ Z_{\rm  tree}^{(3)}=0$, this is equivalent to the one, (\ref{FermiTreeMethodI}), in Method(I). The $1$-loop part reads
\begin{eqnarray}
Z_{1-\mbox{loop}}^{(\hat{i})} = {\cal P}^{(\hat{i})}_0 {\cal L}^{(\hat{i})}_0 \Bigg|_{\Omega_1^{(\hat{i})} \mapsto 0 } \hspace{-3ex}=    \epsilon \! \left( \Omega_0^{(\hat{i})}  \right) 
 \exp \left[ - \frac{ N   \left( \Omega_0^{(\hat{i})} - \omega  \right)^2  }{ 2 \lambda^2 }  \right] 
 \frac{  \left(  \Omega_0^{(\hat{i})} \right)^{N+1}   \left(  \Omega_0^{(\hat{i})} +  \delta \omega  \right) }{ \sqrt{ \left( \Omega_0^{(\hat{i})}  \right)^2 + \lambda^2 }}  \ ,  \label{1-loopZ}  
\end{eqnarray}
and $Z_{1-\mbox{loop}}^{(3)} =0 $ with the aid of (\ref{P_0}) and (\ref{L_0 i=1,2}), which again matches with (\ref{Fermi1-loopMethodI}). Next
\begin{eqnarray}
& &\hspace{-8ex} \left. Z_{2-\mbox{loop}}^{(\hat{i})} ={\cal P}^{(\hat{i})}_0 \left[1 + \frac{ \left({\cal P}^{(\hat{i})}_1 + {\cal L}^{(\hat{i})}_1  \right) }{N} \right] \right|_{\Omega_2^{(\hat{i})} \mapsto 0 } \hspace{-3ex}=  Z_{1-\mbox{loop}}^{(\hat{i})}  \left[  1 +   \frac{  \left({\cal P}^{(\hat{i})}_1 + {\cal L}^{(\hat{i})}_1  \right) }{N}    \right]_{\Omega_2^{(\hat{i})} \mapsto 0 }   \hspace{-3ex} ;   \label{2-loopZ}
\end{eqnarray}
where
\begin{eqnarray}
& &\hspace{-8ex}\left( {\cal P}^{(\hat{i})}_1 + {\cal L}^{(\hat{i})}_1 \right) \Big|_{\Omega_2^{(\hat{i})} \mapsto 0} \hspace{-1ex} = \left( \sum_{r=1}^3 F_1^{(i)}(r) +   \frac{ \Omega_1^{(\hat{i})} }{ \Omega_0^{(\hat{i})} + \delta \omega } +{\cal L}^{(\hat{i})}_1  \right)  \Big|_{\Omega_2^{(\hat{i})} \mapsto 0} \nonumber \\
& & \hspace{4ex} =  - \frac{\left(  \Omega_1^{(\hat{i})} \right)^2 }{2 \lambda^2 } - \frac{1}{2}\left( \frac{  \Omega_1^{(\hat{i})}  }{  \Omega_0^{(\hat{i})}  }\right)^2 -  \frac{ \lambda^2 }{  \left( \Omega_0^{(\hat{i})} \right)^2  +  \lambda^2  } \left[ \frac{  \Omega_1^{(\hat{i})}  }{  \Omega_0^{(\hat{i})}  }- \frac{1}{2} \left( \frac{\Omega_0^{(\hat{i})}}{ \Omega_0^{(\hat{i})} + \delta \omega} \right)^2  \right]  \nonumber \\
& & \hspace{10ex}+   \frac{ \Omega_1^{(\hat{i})} }{ \Omega_0^{(\hat{i})} + \delta \omega }  -  \frac{ 3 \lambda^4 }{4 \left[  \left( \! \Omega_0^{(\hat{i})} \!  \right)^2 \!  \! \!  +  \lambda^2 \!  \right]^2  } +   \frac{ 5 \lambda^6  }{6  \left[ \left( \! \Omega_0^{(\hat{i})} \!  \right)^2 \!  \! \!  +  \lambda^2 \!  \right]^3  } \ , \label{L_1+P_1 2-loop}
\end{eqnarray}
from (\ref{F_1(1) F_2(1)}), (\ref{F_0(2) F_1(2)}), (\ref{F_1(3)}) and (\ref{L_1 i=1,2}). For $i=3$ we obtain
\begin{eqnarray}
& & \hspace{-4ex}Z_{2-\mbox{loop}}^{(3)} = {\cal P}^{(3)}_1 {\cal L}^{(3)}_0 \Bigg|_{2-\mbox{loop}} \nonumber \\
& &  = \epsilon \! \left( \Omega_1^{(3)}  \right)  \exp \left[ - \frac{ N \left( \Omega_0^{(3)} - \omega  \right)^2 }{2 \lambda^2 }  \right]\frac{ \left(  \Omega_0^{(3)} \right)^{N}}{\mathrm e} \frac{\left(  \Omega_1^{(3)} \right)^2 }{ \sqrt{N}\lambda } \frac{1}{N}  \frac{13}{12}  \ ,   \label{Z^3_2-loop}
\end{eqnarray}
in view of (\ref{P_1(3)}) and (\ref{L_0 i=3 2-loop}).

Finally $Z$ under $3$-loop for $i=1,2$ is
\begin{eqnarray}
& &  Z_{3-\mbox{loop}}^{(\hat{i})} = {\cal P}^{(\hat{i})}_0 \left[1 + \frac{{\cal P}^{(\hat{i})}_1 + {\cal L}^{(\hat{i})}_1 }{N} + \frac{ {\cal P}^{(\hat{i})}_2 + {\cal L}^{(\hat{i})}_2 + {\cal P}^{(\hat{i})}_1 {\cal L}^{(\hat{i})}_1}{N^2} \right]  \nonumber \\
& &  = Z_{1-\mbox{loop}}^{(\hat{i})} \left[1 + \frac{{\cal P}^{(\hat{i})}_1 + {\cal L}^{(\hat{i})}_1 }{N} + \frac{ {\cal P}^{(\hat{i})}_2 + {\cal L}^{(\hat{i})}_2 + {\cal P}^{(\hat{i})}_1 {\cal L}^{(\hat{i})}_1}{N^2} \right]  \ ,\label{Z3loop i=1,2} 
\end{eqnarray}
whose functions ${\cal P}^{(\hat{i})}_{1,2}, {\cal L}^{(\hat{i})}_{1,2}$ are given in the appendix~\ref{sec:appendixC}; (\ref{P_1^hati}), (\ref{P_2^hati}), (\ref{L_1 i=1,2}), and (\ref{L_2 i=1,2}). 

Meanwhile $i=3$ is
\begin{eqnarray}
& & \hspace{-3ex} Z_{3-\mbox{loop}}^{(3)} \hspace{-1ex}=\frac{{ \cal P}^{(3)}_1}{N} \left( 1 + \frac{{ \cal P}^{(3)}_2}{N} \right) \left( { \cal L}^{(3)}_0 + \frac{{ \cal L}^{(3)}_1}{N} \right) \Bigg|_{3-\mbox{loop}} \hspace{-4ex}= \frac{{ \cal P}^{(3)}_1}{N} \left[ { \cal L}^{(3)}_0 + \frac{ { \cal L}^{(3)}_1 + { \cal L}^{(3)}_0 { \cal P}^{(3)}_2 }{N}  \right] \Bigg|_{3-\mbox{loop}} \nonumber \\
& & \hspace{5ex}   = \epsilon \! \left( \Omega_1^{(3)}  \right)   \exp \left[ - \frac{ N \left( \Omega_0^{(3)} - \omega  \right)^2 }{2 \lambda^2 }  \right] \frac{\left(  \Omega_0^{(3)} \right)^{N} \left(  \Omega_1^{(3)} \right)^2}{{\mathrm e} \sqrt{N}\lambda }  \nonumber \\
& &  \hspace{8ex} \times  \left[   \frac{1}{N} \frac{313}{288}   +  \frac{1}{N^2} 
\left(  - \frac{13}{12} \left[   \left( \frac{ \Omega _1^{(3)} }{ \Omega_0^{(3)} }  \right)^2 +   \frac{ \left( \Omega_1^{(3)} \right)^2  }{ \lambda^2 } \right]   +   \frac{313}{288} { \cal P}^{(3)}_2    \right)   \right]  \ , \label{Z^3_3-loop}
\end{eqnarray}
where ${ \cal P}^{(3)}_2$ is given in (\ref{P_2^3}) and use has been made of (\ref{L^3_0 3-loop}) and (\ref{L^3_1 Result}).
For $Z_{\mbox{tree}}$ and $ Z_{1-\mbox{loop}}$, there is no difference from Method (I). However, in $2$- and $3$-loop, the third saddle starts contributing to give
\begin{eqnarray}
Z^{\rm Total}_{l-\mbox{loop}} = Z^{(1)}_{l-\mbox{loop}} + Z^{(2)}_{l-\mbox{loop}} + Z^{(3)}_{l-\mbox{loop}} \ ; \quad (l=2,3) \ , 
\end{eqnarray}
whose numerical results, when $N=2$ and $\omega=1$ with $\omega_0 = 10^{2}$ and $1$, are equivalent to those of Method(I) for $10^{-3} \leq \lambda \leq 10$, which is in the table~\ref{table: FermiMethod(I)}. In the table~\ref{table: FermiMethod(II)}, we list the result of $\omega_0 = 10^{-2}$, in which we see disparities at $2$- and $3$-loop in the weak coupling region $10^{-2} \leq \lambda \leq 1$. Discrepancies are notable, reaching to $\sim 600$ times to the exact value at $\lambda= 10^{-1}$ in $3$-loop.

\begin{table}[h]
{\tiny 
\begin{eqnarray*}
\begin{array}{|l|cccccc|}
\hline\hline
 \omega_0 = 10^{-2}
 &\lambda & \textrm{Exact} &
 \begin{array}{c} \textrm{ tree}\\ \textrm{ (tree)/Exact}\\ \end{array} &
 \begin{array}{c} \textrm{ 1-loop}\\ \textrm{ (1-loop)/Exact}\\ \end{array} &
 \begin{array}{c} \textrm{ 2-loop}\\ \textrm{ (2-loop)/Exact}\\ \end{array} &
 \begin{array}{c} \textrm{ 3-loop}\\ \textrm{ (3-loop)/Exact}\\ \end{array} \\
\cline{2-7}
 &10^{-3} & 1.0001\times 10^{-2} &
  \begin{array}{c} 1.0001\times 10^{-2}\\ 1.0000\\ \end{array} &
  \begin{array}{c} 1.0001\times 10^{-2}\\ 1.0000\\ \end{array} &
  \begin{array}{c} 1.0001\times 10^{-2}\\ 1.0000\\ \end{array} &
  \begin{array}{c} 1.0001\times 10^{-2}\\ 1.0000\\ \end{array} \\
\gcline{2-7}
 &10^{-2} & 1.0101\times 10^{-2} &
  \begin{array}{c} 1.0101\times 10^{-2}\\ 1.0000\\ \end{array} &
  \begin{array}{c} 1.0100\times 10^{-2}\\ 1.0000\\ \end{array} &
  \begin{array}{c} 9.6026\times 10^{-3}\\ 0.9507\\ \end{array} &
  \begin{array}{c} 1.0335\times 10^{-2}\\ 1.0232\\ \end{array} \\
\gcline{2-7}
 &10^{-1} & 2.0050\times 10^{-2} &
  \begin{array}{c} 2.0100\times 10^{-2}\\ 1.0025\\ \end{array} &
  \begin{array}{c} 2.0002\times 10^{-2}\\ 0.9976\\ \end{array} &
  \begin{array}{c} -3.1834\times 10^{-1}\\-15.877\\ \end{array} &
  \begin{array}{c} 1.2398\times 10^{1}\\  618.36\\ \end{array} \\
\gcline{2-7}
 &1       & 1.0150 &
  \begin{array}{c} 1.0774\\ 1.0615\\ \end{array} &
  \begin{array}{c} 0.9311\\ 0.9173\\ \end{array} &
  \begin{array}{c} 0.8719\\ 0.8590\\ \end{array} &
  \begin{array}{c} 1.2693\\ 1.2506\\ \end{array} \\
\gcline{2-7}
 &10      & 1.0051\times 10^{2} &
  \begin{array}{c} 1.1105\times 10^{2}\\  1.1049\\ \end{array} &
  \begin{array}{c} 9.1577\times 10^{1}\\  0.9111\\ \end{array} &
  \begin{array}{c} 9.9931\times 10^{1}\\  0.9942\\ \end{array} &
  \begin{array}{c} 1.0085\times 10^{2}\\  1.0034\\ \end{array} \\
\hline\hline
\end{array}
\end{eqnarray*}

\vspace{-5ex}

\caption{Result of Method(II): $N=2$, $\omega=1$ and $\omega_0 = 10^{-2}$ for $10^{-3} \leq \lambda \leq 10$. Discrepancies to Method(I), in the table~\ref{table: FermiMethod(I)}, are acknowledged in $2,3$-loop for $10^{-2} \leq \lambda \leq 10$.} 
\label{table: FermiMethod(II)}
}
\end{table}

\begin{figure}[h]
$$
\includegraphics[width=7cm]{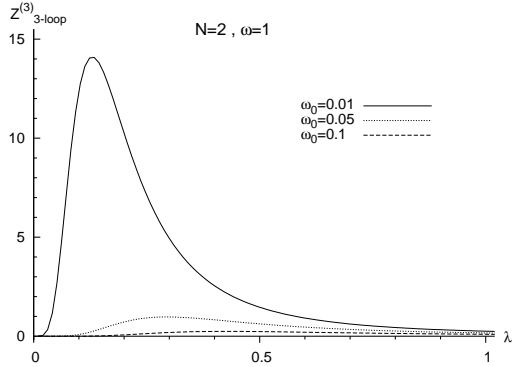} 
$$

\vspace{-3ex}

\caption{The graph of $Z_{3-\mbox{loop}}^{(3)}$ is shown with $N=2, \omega=1$ for $0 \leq \lambda \leq 1$. The vertical line is the value of $Z_{3-\mbox{loop}}^{(3)}$ itself. The solid, dotted, and dashed line designate $\omega_0 =0.01, 0.05, 0.1$. The deviation from the exact value (which is almost zero) is seen around $\lambda \sim 1.5$ when $\omega_0 = 0.01$.}\label{Fig:Z^3_3-loop}
\end{figure}

The reason can be seen from $Z_{3-\mbox{loop}}^{(3)}$(\ref{Z^3_2-loop}) by putting $\omega_0 \ll 1$; 
\begin{eqnarray}
Z_{3-\mbox{loop}}^{(3)} \sim \exp \left[ -N \frac{\omega_0^2 }{2 \lambda^2 } \right] \frac{1}{\lambda } \ ; 
\end{eqnarray}
which has some peak around $\lambda \sim \omega_0$. The graph is shown in the figure~\ref{Fig:Z^3_3-loop}, implying a large deviation at the value $\omega_0 = 0.01$. 
Although the case would not be included in the real physical situation, u- and d-quarks are lighter than s-quark, Method(II) is worse than (I) in the above situation. Therefore, a recipe for an approximation Method(I) is better than (II) and moreover simpler for actual calculations.

\clearpage
\section{Discussion}\label{sec:discussion}
Under Method(II), we encounter the same situation in the bosonic four-body model under the weak coupling region when the "mass", $\omega_0$, is tiny: consider $\mbit{\sigma}$ (\ref{Model1}) in the introduction as bosonic variables, to have
\begin{eqnarray}
Z^{\mbit{\omega }}\sim  \int   \! \!  \frac{dy}{\omega_0 + i \lambda y} \exp \left[ -Nf(y) \right]  \ , 
\end{eqnarray}
where in view of (\ref{E(y)}) and (\ref{f(y)Def.}),
\begin{eqnarray}
f(y) =   \frac{y^2}{2} +  \ln \left( \omega + i \lambda y \right)    \ . 
\end{eqnarray}
(We have omitted irrelevant factors, $2\pi , N$.) Therefore
\begin{eqnarray}
& & Z^{\mbit{\omega }}\sim  \int \! \! dt \exp \left[ -N \tilde{f}(t) \right] \ ,   \\
& & \hspace{-5ex}\tilde{f}(t) \equiv f(t) + \frac{1}{N}\ln g(t)  \ ;  \quad g(t) \equiv   \omega_0 + i \lambda t  \ . \label{Bosnic g(t)}
\end{eqnarray}
(Again we have switched, $y \mapsto t$.) 

Take, for the time being, $\tilde{f}(t), f(t), g(t)$ as generic, in other words, start from (\ref{Method2}) to make a general discussion:
there emerge additional saddle points, $t^A_0$, in the gap equation when $N \mapsto \infty$,
\begin{eqnarray}
 0 = f'(t) + \frac{1}{N}\frac{g'(t)}{g(t)} \stackrel{N \mapsto \infty}{\Longrightarrow} g(t)f'(t)  = 0 \ ; 
\end{eqnarray}
other than $f'(t_0)=0$ such that
\begin{eqnarray}
g \! \left( t^A_0 \right)  = 0 \ .
\end{eqnarray}
Therefore the additional saddle point(s) is expanded as
\begin{eqnarray}
t_c = t^A_0 + \frac{t^A_1}{N } + O \! \left( \frac{1}{N^2} \right)  \ ,
\end{eqnarray}
around which
\begin{eqnarray}
g_c \equiv g (t_c) = \frac{t^A_1}{N}  g^{(1)}_A  +  O \! \left( \frac{1}{N^2} \right) \ ; \quad  g^{(1)}_A \equiv g^{(1)} \! \left( t^A_0 \right) \ .  \label{g_cExpansion}
\end{eqnarray}
Now recall that one of the $2$-loop terms is given as ( (\ref{I_NTildeFinal}) with (\ref{TildeF Definition}) )
\begin{eqnarray}
 Z^{\mbit{\omega }}_{2-\mbox{loop}} \sim \frac{1}{\sqrt{N \tilde{f}^{(2)}_c}} \exp \left[ -N \tilde{f}_c \right] \frac{\tilde{f}^{(4)}_c}{N  \tilde{f}^{(2)}_c }= \frac{\tilde{f}^{(4)}_c}{g_c \left( N \tilde{f}^{(2)}_c \right)^{3/2} } \exp \left[ -N f_c \right] \ . 
\end{eqnarray}
By noting
\begin{eqnarray}
\tilde{f}^{(n)}_c = f^{(n)}_c + \frac{1}{N}\left[ (-)^{n-1} n!\left(  \frac{g_c^{(1)}}{g_c} \right)^n + O \! \left( \frac{1}{\left( g_c \right)^{n-1} } \right)  \right] \   , 
\end{eqnarray}
we find
\begin{eqnarray}
 Z^{\mbit{\omega }}_{2-\mbox{loop}} \sim  \frac{\tilde{f}^{(4)}_c}{g_c \left( N \tilde{f}^{(2)}_c \right)^{3/2} } \exp \left[ -N f_c \right] \sim \frac{g_c^{(1)}}{(g_c)^2} \exp \left[ -N f_c \right] \ , \label{2-loopZaroundNEWSADDLE}
\end{eqnarray}
around the additional saddle point, which reads in the bosonic case (\ref{Bosnic g(t)})
\begin{eqnarray}
 Z^{\mbit{\omega }}_{2-\mbox{loop}} \sim   \frac{g_c^{(1)}}{(g_c)^2} \exp \left[ -N f_c \right] \stackrel{\omega_0 \mapsto 0 }{\sim } \frac{1}{\lambda }  \ , 
\end{eqnarray}
since $g_c \sim  g_c^{(1)} \sim \lambda $ (\ref{Bosnic g(t)}), again implying a large deviation. 

In summary, Method(II) was superficially simpler than (I) but needs a rather cumbersome procedure in the actual calculation and moreover always seems to suffer from a large deviation when in a weak coupling region when the one "mass" $\omega_0$ is tiny.

As the final comment, we check the validity of Method(I) in an alternative way: the case of $\omega_0 = \omega$ corresponds to the $N=3$ version of the model\cite{rf:kashiSaka},
\begin{eqnarray}
Z_{\rm KS} \equiv \int d^{N}\hat{\mbit{\xi}}d^{N}\hat{\mbit{\xi}}^* \exp \left[ -\hat{\mbit{\xi}}^*\cdot\mbit{\omega} \cdot \hat{\mbit{\xi}} + \frac{\left( \lambda_{\rm KS} \right) ^2}{2 N} \left( \hat{\mbit{\xi}}^* \cdot \hat{\mbit{\xi}}\right)^2 \right]  \ . \label{Z_KS}
\end{eqnarray}
So if we put
\begin{eqnarray}
\left( \lambda_{\rm KS} \right) ^2 = \frac{3}{2} \lambda^2 \ ,
\end{eqnarray}
this should agree with our model($N=2$). Applying a usual AFM (that is, from the expression (\ref{Model}) to (\ref{stabilityCond})), we have results up to $3$-loop, which is listed in the table \ref{table:KS_N3}. By comparing this with the table \ref{table: FermiMethod(I)} (of $\omega_0 = \omega$), there is no big difference: almost all data show that the standard treatment of the table \ref{table:KS_N3} yields a slightly better value except the tree approximation in $ 10^{-1} \leq \lambda \leq 10$ where our model under Method(I) results better.   
\begin{table}[h]
{\tiny
$$
\begin{array}{|cccccc|}
\hline\hline
 \lambda & \textrm{Exact} &
 \begin{array}{c} \textrm{ tree}\\ \textrm{ (tree)/Exact}\\ \end{array} &
 \begin{array}{c} \textrm{ 1-loop}\\ \textrm{ (1-loop)/Exact}\\ \end{array} &
 \begin{array}{c} \textrm{ 2-loop}\\ \textrm{ (2-loop)/Exact}\\ \end{array} &
 \begin{array}{c} \textrm{ 3-loop}\\ \textrm{ (3-loop)/Exact}\\ \end{array} \\
\hline
 \sqrt{\frac{3}{2}}\times 10^{-3} & 1.0000 &
  \begin{array}{c} 1.0000 \\ 1.0000 \\ \end{array} &
  \begin{array}{c} 1.0000 \\ 1.0000 \\ \end{array} &
  \begin{array}{c} 1.0000 \\ 1.0000 \\ \end{array} &
  \begin{array}{c} 1.0000 \\ 1.0000 \\ \end{array} \\
\ghline
 \sqrt{\frac{3}{2}}\times 10^{-2} & 1.0002 &
  \begin{array}{c} 1.0002 \\ 1.0001 \\ \end{array} &
  \begin{array}{c} 1.0002 \\ 1.0000 \\ \end{array} &
  \begin{array}{c} 1.0002 \\ 1.0000 \\ \end{array} &
  \begin{array}{c} 1.0002 \\ 1.0000 \\ \end{array} \\
\ghline
 \sqrt{\frac{3}{2}}\times 10^{-1} & 1.0150 &
  \begin{array}{c} 1.0224 \\ 1.0073 \\ \end{array} &
  \begin{array}{c} 1.0151 \\ 1.0001 \\ \end{array} &
  \begin{array}{c} 1.0150 \\ 1.0000 \\ \end{array} &
  \begin{array}{c} 1.0150 \\ 1.0000 \\ \end{array} \\
\ghline
 \sqrt{\frac{3}{2}} & 2.5000 &
  \begin{array}{c} 3.0574 \\ 1.2230 \\ \end{array} &
  \begin{array}{c} 2.5433 \\ 1.0173 \\ \end{array} &
  \begin{array}{c} 2.5032 \\ 1.0013 \\ \end{array} &
  \begin{array}{c} 2.4996 \\ 0.9998 \\ \end{array} \\
\ghline
 \sqrt{\frac{3}{2}}\times 10 & 1.5100\times 10^{2} &
  \begin{array}{c} 2.0188\times 10^{2} \\ 1.3369 \\ \end{array} &
  \begin{array}{c} 1.5484\times 10^{2} \\ 1.0254 \\ \end{array} &
  \begin{array}{c} 1.5106\times 10^{2} \\ 1.0004 \\ \end{array} &
  \begin{array}{c} 1.5089\times 10^{2} \\ 0.9993 \\ \end{array} \\
\hline\hline
\end{array}
$$

\vspace{-4ex}

\caption{Results of the model (\ref{Z_KS}) at $N=3$, $\omega=1$, and $\lambda_{\rm KS}=\sqrt{\frac{3}{2}} \lambda $. }\label{table:KS_N3}
}
\end{table}
\vspace{10mm}

\noindent{\Large \bf Acknowledgements}

\vspace{5mm}

The authors are grateful to H. So for discussions, especially for guiding them to the second proof of (\ref{no1/Ncontribution}).

\appendix

\section{The table of $T \! \left( L, k | 2L-k \right) $ and $T \! \left( L, k | 2L-k -1\right)$ defined by eq.(\ref{T(L,k|  )Def.Q})}\label{sec:appendixA}
In this appendix, we list the table of  $T \! \left( L, k | 2L-k \right)/ \left(  2(L+k)-1  \right)!! $ instead of  $T \! \left( L, k | 2L-k \right) $ itself; since otherwise the values become very large. The range up to $L=14$ is needed for the calculation of the Gamma function in Sec.\ref{sec:Gamma} and to $L=4$ in Sec.\ref{sec:Fermion}.
\begin{table}[h]

\tiny{
$$
\begin{array}{
|@{\hspace{0.2em}}r@{\hspace{0.2em}}
|@{\hspace{0.2em}}r@{\hspace{0.2em}}
|@{\hspace{0.2em}}r@{\hspace{0.2em}}
|@{\hspace{0.2em}}r@{\hspace{0.2em}}
|@{\hspace{0.2em}}r@{\hspace{0.2em}}
|@{\hspace{0.2em}}r@{\hspace{0.2em}}
|@{\hspace{0.2em}}r@{\hspace{0.2em}}
|@{\hspace{0.2em}}r@{\hspace{0.2em}}
|@{\hspace{0.2em}}r@{\hspace{0.2em}}|}\hline
k\backslash L&0&1&2&3&4&5&6&7\\\hline
0&1&0&0&0&0&0&0&0\\\ghline
1&&-1/4&-1/6&-1/8&-1/10&-1/12&-1/14&-1/16\\\ghline
2&&1/18&47/480&153/1400&3349/30240&42131/388080&605453/5765760&655217/6486480\\\ghline
3&&&-1/72&-493/17280&-4049/100800&-59197/1209600&-161453/2910600&-81158813/1345344000\\\ghline
4&&&1/1944&77/25920&190261/29030400&1722811/163296000&406957909/27941760000&668430857/36324288000\\\ghline
5&&&&-1/7776&-503/933120&-143921/116121600&-5689699/2612736000&-39726066467/12070840320000\\\ghline
6&&&&1/524880&107/4665600&44021/522547200&74863031/376233984000&4489161401/12070840320000\\\ghline
7&&&&&-1/2099520&-1699/503884800&-216793/18811699200&-13916027/501645312000\\\ghline
8&&&&&1/264539520&137/1763596800&282133/658409472000&142181689/101583175680000\\\ghline
9&&&&&&-1/1058158080&-643/63489484800&-1143257/23702740992000\\\ghline
10&&&&&&1/214277011200&167/1142810726400&361657/319987003392000\\\ghline
11&&&&&&&-1/857108044800&-29/1645647446016\\\ghline
12&&&&&&&1/254561089305600&197/1131382619136000\\\ghline
13&&&&&&&&-1/1018244357222400\\\ghline
14&&&&&&&&1/416971064282572800\\\hline
\end{array}
$$\\

\vspace{-8ex}

$$
\begin{array}{
|@{\hspace{0.2em}}r@{\hspace{0.2em}}
|@{\hspace{0.2em}}r@{\hspace{0.2em}}
|@{\hspace{0.2em}}r@{\hspace{0.2em}}
|@{\hspace{0.2em}}r@{\hspace{0.2em}}|}\hline
k\backslash L&8&9&10\\\hline
0&0&0&0\\\ghline
1&-1/18&-1/20&-1/22\\\ghline
2&23763863/245044800&158899519/1707145440&11098301/124156032\\\ghline
3&-232229821/3632428800&-1974182737/29640619008&-176989210093/2581203905280\\\ghline
4&3218616617/146459529216&131020195003/5187108326400&1429205465892419/50591596543488000\\\ghline
5&-59153229587/13076743680000&-237186996829/40683202560000&-4136077769339/576345369600000\\\ghline
6&449840404627/747242496000000&2187240114496949/2471504555520000000&27225846275010179/22408307970048000000\\\ghline
7&-15727453241/289700167680000&-11643095464009/125536739328000000&-711993190614367/4943009111040000000\\\ghline
8&570158851513/166867296583680000&23485635969143/3389491961856000000&4800125589203449/388250897448960000000\\\ghline
9&-434386633/2844328919040000&-754670748947/2002407559004160000&-28426615868773/36154580926464000000\\\ghline
10&1125369389/230390642442240000&4768426956641/315379190543155200000&61720076581079809/1639971790824407040000000\\\ghline
11&-142837/1279948013568000&-414828299/921562569768960000&-46647858451427/34060952578660761600000\\\ghline
12&2278697/1267148533432320000&64781047/6516763886223360000&14267906296747/374670478365268377600000\\\ghline
13&-347/17455617552384000&-7370813/45617347203563520000&-147673007/182469388814254080000\\\ghline
14&227/1588461197266944000&563401/296512756823162880000&8377793029/640467554738031820800000\\\ghline
15&-1/1667884257130291200&-6271/400292221711269888000&-1697149/10674459245633863680000\\\ghline
16&1/900657498850357248000&257/3002191662834524160000&957911/672490932474933411840000\\\ghline
17&&-1/3602629995401428992000&-983/108078899862042869760000\\\ghline
18&&1/2480410751833883860992000&41/1049909312945559306240000\\\ghline
19&&&-1/9921643007335535443968000\\\ghline
20&&&1/8483004771271882804592640000\\\hline
\end{array}
$$
}

\vspace{-5ex}

\caption{The values of $T(L,k|2L-k)/ \left(  2(L+k)-1  \right)!!$ for $L=0 \sim 10$ in sec.\ref{sec:Gamma} as well as sec.\ref{sec:Fermion}}\label{table:=2L-k:1}
\end{table}
\begin{table}[h]
\tiny{
$$
\begin{array}{
|@{\hspace{0.2em}}r@{\hspace{0.2em}}
|@{\hspace{0.2em}}r@{\hspace{0.2em}}
|@{\hspace{0.2em}}r@{\hspace{0.2em}}|}\hline
k\backslash L&11&12\\\hline
0&0&0\\\ghline
1&-1/24&-1/26\\\ghline
2&265842403/3093554464&20666950267/249864014400\\\ghline
3&-4160074481/59435616240&-205489485020713/2894174878795200\\\ghline
4&2665668285176389/86090742017280000&15599534488926830887/466772524452089856000\\\ghline
5&-8637934672612781/1011831930869760000&-40541391139263449/4097919320022528000\\\ghline
6&203058721852862831/128059978750704000000&40047282902251643797/20117058480110592000000\\\ghline
7&-841208011558860343/4033495434608640000000&-93845936786285052059/327833545601802240000000\\\ghline
8&405347221415512403/20167477173043200000000&3993321076456790080543/131133418240720896000000000\\\ghline
9&-198721338986302927/136664315902033920000000&-3058025156141685767/1244621448393523200000000\\\ghline
10&691811729911733023/8609851901828136960000000&21345802689252486829/139397602220074598400000000\\\ghline
11&-202677271543968731/59038984469678653440000000&-1543928961441550217/206636445643875287040000000\\\ghline
12&3374227932094529/29519492234839326720000000&19789232975365781149/68666880398580095385600000000\\\ghline
13&-40177324205483/13488137221149661593600000&-103383349078912121/11689718924996373381120000000\\\ghline
14&79944551761007/1315093379062092005376000000&296393081576585261/1367697114224575685591040000000\\\ghline
15&-2481097459/2561870218952127283200000&-22322956680727/5260373516248368021504000000\\\ghline
16&2893597613/242096735690976028262400000&14641217114597/220935687682431456903168000000\\\ghline
17&-101081/896654576633244549120000&-794890043/968386942763904113049600000\\\ghline
18&74363/93537375153331647283200000&5917683217/740816011214386646482944000000\\\ghline
19&-9637/2381194321760528506552320000&-6685391/111122401682157996972441600000\\\ghline
20&317/22621346056725020812247040000&56741/165594167212627603331481600000\\\ghline
21&-1/33932019085087531218370560000&-61/42861497791689513117941760000\\\ghline
22&1/35272333838948488701496197120000&347/85508688094420578670293811200000\\\ghline
23&&-1/141089335355793954805984788480000\\\ghline
24&&1/175232954511896091869033107292160000\\\hline
\end{array}
$$\\

\vspace{-12ex}

$$
\begin{array}{
|@{\hspace{0.2em}}r@{\hspace{0.2em}}
|@{\hspace{0.2em}}r@{\hspace{0.2em}}
|@{\hspace{0.2em}}r@{\hspace{0.2em}}|}\hline
k\backslash L&13&14\\\hline
0&0&0\\\ghline
1&-1/28&-1/30\\\ghline
2&192066102203/2409402996000&5733412167187/74530866009600\\\ghline
3&-316144708749043/4410171243878400&-5230982263631/72535711248000\\\ghline
4&1039787017585599191/29173282778255616000&86993393978615452781/2310523996037844787200\\\ghline
5&-1321164558963178417729/117626676161926643712000&-409853074642336437191/32674076711646289920000\\\ghline
6&124759585247032919932373/51461670820842906624000000&515724686858368491454901/178997115898584023040000000\\\ghline
7&-3511001126573317919533/9317374453945958400000000&-26661050735382530348701583/55578604486510339153920000000\\\ghline
8&9940709833698231428297/227610147374965555200000000&15680328374701897384850153/261546374054166301900800000000\\\ghline
9&-41120299919631976751017/10621806877498392576000000000&-91936525080313650485417/15932710316247588864000000000\\\ghline
10&136358632264218467927423/509846730119922843648000000000&12636723159765370307618303/28980761501553509007360000000000\\\ghline
11&-1026261079863684476899/70256391518917597593600000000&-40157529586233540547571/1529540190359768530944000000000\\\ghline
12&2015526363139860379999/3161537618351291891712000000000&23571578581358615769453103/18501318142591760150298624000000000\\\ghline
13&-880105003537523330687/39277455587987814560563200000000&-83901448111763933230733/1669291862489482118823936000000000\\\ghline
14&338122116167654792537/530245650437835496567603200000000&884333215764851072621/545074893874116610228224000000000\\\ghline
15&-26876946960034477/1823596152299434247454720000000&-1186820857702705430201/27572773822767445821515366400000000\\\ghline
16&1032930932083247/3730083038794297324339200000000&1167709777320002702963/1240774822024535061968191488000000000\\\ghline
17&-302359614816397/71583162809107792036626432000000&-526519789986178363/31019370550613376549204787200000000\\\ghline
18&31701146999281/608456883877416232311324672000000&266111014317840767/1054658598720854802672962764800000000\\\ghline
19&-4588191221/8889792134572639757795328000000&-40513735354423/13142668691752190617924612915200000\\\ghline
20&9262374577/2280231682517882097874501632000000&576706185137543/18728302885746871630542573404160000000\\\ghline
21&-7598699/304030891002384279716600217600000&-2275833187/9120926730071528391498006528000000\\\ghline
22&5619821/47884865332875524055364534272000000&1596747533/985060086847725066281784705024000000\\\ghline
23&-13723/33861440485390549153436349235200000&-17489411/2106934074646523058436039507968000000\\\ghline
24&29/29954351198614716558809078169600000&14256551/436135353451830273096260178149376000000\\\ghline
25&-1/700931818047584367476132429168640000&-211/2213468899097634844661470828953600000\\\ghline
26&1/1025112783894592137433843677659136000000&37/191163223103886645675308844318720000000\\\ghline
27&&-1/4100451135578368549735374710636544000000\\\ghline
28&&1/6974867381618804903099872382792761344000000\\\hline
\end{array}
$$
}

\vspace{-5ex}

\caption{The values of $T(L,k|2L-k)/ \left(  2(L+k)-1  \right)!!$ for $L=11 \sim 14$ in sec.\ref{sec:Gamma} as well as sec.\ref{sec:Fermion}}\label{table:=2L-k:2}

\end{table}

\section{The proof of the relation (\ref{Equality=InEqality}) and of (\ref{no1/Ncontribution})}\label{sec:appendixB}

In this appendix, we first prove (\ref{Equality=InEqality}): 
\begin{eqnarray}
 \sum_{k=0}^{2L} T \! \left(L, k \ \big| \leq 2L-k  \right)=  \sum_{k=0}^{2L} T \! \left( L,k \ \big| 2L-k  \right)  \ , \label{Target0}
\end{eqnarray}
where 
\begin{eqnarray}
& & T \! \left(L, k  \left| \leq 2L-k \right. \right)  =  (-)^k  \left( 2(L+k) -1\right)!! \  \sum_{\mbox{all possible} \ \{ A_\alpha \} \hfill \atop
\scriptstyle  \sum_{\alpha=1 }^P Q_\alpha= k \  ; \  \left(  P \leq k  \right) }^{ \sum_{\alpha=1 }^P Q_\alpha A_\alpha  \leq  2L-k }   \hspace{-2ex} \frac{1}{Q_1!\cdots Q_P!}  \nonumber \\ 
& & \hspace{28ex} \times      \frac{1}{\left( A_1 + 3 \right)^{Q_1} \left( A_2 + 3 \right)^{Q_2} \cdots \left( A_P + 3 \right)^{Q_P}} \  ,  \label{T(L,k| < )Def.Q in Apendix}
\end{eqnarray}
and $T \! \left( L,k \ \big| 2L-k  \right)$ in RHS is given in terms of the conditional sum $\sum_{\alpha=1 }^P Q_\alpha A_\alpha  =  2L-k$ instead of $\sum_{\alpha=1 }^P Q_\alpha A_\alpha  \leq  2L-k$. When $L=0$ (which also means $ k=0$) the relation (\ref{Target0}) holds
\begin{eqnarray}
T \! \left(0, 0 \ \big| \leq 0  \right) = T \! \left(0, 0 \ \big|  0  \right) \ ,
\end{eqnarray}
since $\sum_{\alpha =1}^P Q_\alpha A_\alpha \leq 0 $ is nothing but $\sum_{\alpha =1}^P Q_\alpha A_\alpha = 0 $ for any positive $Q_\alpha , A_\alpha$. Therefore we can set $L\neq 0$ in the following. Likewise when $k=2L$
\begin{eqnarray}
T \! \left(L, 2L \ \big| \leq 0  \right) = T \! \left(L, 2L \ \big|  0  \right) \ . 
\end{eqnarray}
Therefore the target relation (\ref{Target0}) turns out to be
\begin{eqnarray}
 \sum_{k=0}^{2L-1} T \! \left(L, k \ \big| < 2L-k  \right)= 0  \ , \qquad L \neq 0 \ ,  \label{Target1} 
\end{eqnarray}
with $ T \! \left(L, k \ \big| < 2L-k  \right)$ being given by the conditional sum $\sum_{\alpha=1 }^P Q_\alpha A_\alpha  <  2L-k$ in (\ref{T(L,k| < )Def.Q in Apendix}), which is further rewritten as
\begin{eqnarray}
& & \mbox{LHS of (\ref{Target1})}  =  T \! \left(L, 0 \ \big|< 2L  \right) + T \! \left( L,1 \ \big|2L-2  \right) \nonumber \\ 
& & \hspace{3ex}+ \sum_{k=1}^{2L-2} \Big\{ T \! \left(L, k \ \big|\leq 2L-k-2  \right) + T \! \left(L, k+1 \ \big|2L-k-2  \right) \Big\}   \ ,       \label{TargetFinal}
\end{eqnarray}
since with the aid of (an obvious relation)
\begin{eqnarray}
T \! \left( L, k \ \big| < 2L-k  \right)= T \! \left( L, k \ \big|2L-k-1  \right) + T \! \left(L , k \ \big| \leq  2L-k-2  \right) \ ,
\end{eqnarray}
LHS reads
\begin{eqnarray}
& & \hspace{-4ex}\mbox{LHS of (\ref{Target1})}=  T \! \left( L, 0 \ \big|< 2L  \right) + \sum_{k=1}^{2L-1} \left\{ T \! \left( L, k \ \big|2L-k-1  \right) + T \! \left(L , k \ \big| \leq  2L-k-2  \right) \right\}  \nonumber \\
& &  \hspace{10ex}=  T \! \left( L, 0 \ \big|< 2L  \right) + T \! \left( L, 1 \ \big|2L-2  \right) + \sum_{k=2}^{2L-1}T \! \left( L, k \ \big|2L-k-1  \right)  \nonumber \\
& & \hspace{47ex} +  \sum_{k=1}^{2L-1} T \! \left(L , k \ \big| \leq  2L-k-2  \right) \ , \label{to targetFinal} 
\end{eqnarray}
whose last term becomes $ \sum_{k=2}^{2L-2}T \! \left( L, k \ \big| \leq 2L-k-2  \right)$, because $T \! \left( L, 2L-1 \ \big|\leq -1  \right)=0$, giving the third term of (\ref{TargetFinal}). Meanwhile the third term in (\ref{to targetFinal}) gives the last term of (\ref{TargetFinal}) by shifting $k \mapsto k+1$.

The first term cancels the second in RHS of (\ref{TargetFinal}); since from (\ref{T(L,k| < )Def.Q in Apendix})
\begin{eqnarray}
 T \! \left(L, 0 \ \big|< 2L  \right)= (2L-1)!!    \ ,
\end{eqnarray}
and
\begin{eqnarray}
 T \! \left(L, 1 \ \big|2L-2  \right)= - (2L+1)!! \frac{1}{(2L-2)+3}= - (2L-1)!! \ , 
\end{eqnarray}
from (\ref{T(L,k| < )Def.Q in Apendix}) (with changing the conditional sum to $ Q_1 A_1 = 2L-2 $) and $Q_1= 1$ ( obtaining from $\sum_{\alpha=1}^P Q_\alpha =1 \ ; \ P \leq 1$ ). Therefore if we show 
\begin{eqnarray}
T \! \left(L, K+1 \ \big| 2L-K-2  \right) = - T \! \left(L, K \ \big| \leq 2L-K-2  \right) \ ; \quad  1 \leq K \leq 2L-2 \  . \label{Lemma}
\end{eqnarray}
we find that RHS of (\ref{TargetFinal}) vanishes, accomplishing the proof of (\ref{Target0}).

Now prove (\ref{Lemma}): LHS reads 
\begin{eqnarray}
& & \hspace{-8ex} T \! \left(L, K+1 \ \big| 2L-K-2  \right)= (-)^{K+1} \left( 2(L+K)+1 \right)!!   \nonumber \\ 
& & \hspace{5ex} \times   \sum_{\mbox{all possible} \ \{ A_\alpha \} \hfill \atop
\scriptstyle  \sum_{\alpha=1 }^P Q_\alpha= K+1 \  ; \  \left(  P \leq K+1  \right) }^{ \sum_{\alpha=1 }^P Q_\alpha A_\alpha  = 2L-K-2 } \hspace{0ex} \frac{1}{Q_1!\cdots Q_P!} \ 
\frac{1}{\left( A_1 + 3 \right)^{Q_1}  \cdots \left( A_P + 3 \right)^{Q_P}}    \ . \label{k=K+1}  
\end{eqnarray}
In view of (\ref{T(L,k| < )Def.Q in Apendix}), the conditional sum of RHS, $ \sum_{\alpha =1}^{P} Q_{\alpha} A_{\alpha}  \leq 2L-K-2$, is fulfilled by putting some $ Q_\beta \  ; \ (\beta \in  \alpha )$ to $Q_\beta  -1$, which 
brings the sum $\sum_{\alpha=1 }^P Q_\alpha= K$ to $\sum_{\alpha=1 }^P Q_\alpha -1= K$, while keeping $\sum_{\alpha =1}^{P} Q_{\alpha} A_{\alpha} = 2L-K-2$, to give
\begin{eqnarray}
& &  \hspace{-4ex}  T \! \left(L, K \ \big| \leq 2L-K-2  \right) = (-)^{K} \left( 2(L+K)-1 \right)!! \sum_{\mbox{all possible} \ \{ A_\alpha \} \hfill \atop
\scriptstyle  \sum_{\alpha=1 }^P Q_\alpha= K+1 \  ; \  \left(  P \leq K+1  \right) }^{ \sum_{\alpha=1 }^P Q_\alpha A_\alpha  = 2L-K-2 } \nonumber \\ 
& &  \times   \frac{1}{Q_1!\cdots (Q_\beta -1)! \cdots  Q_P!}   \frac{1}{\left( A_1 + 3 \right)^{Q_1} \cdots  \left( A_\beta  + 3 \right)^{Q_\beta -1} \cdots \left( A_P + 3 \right)^{Q_P}}  \ . \label{RHS of Lemma}
\end{eqnarray}
RHS of (\ref{RHS of Lemma}) is further rewritten as
\begin{eqnarray}
& & \hspace{-4ex}\mbox{RHS of (\ref{RHS of Lemma})} = (-)^{K} \left( 2(L+K)-1 \right)!!  \hspace{-2ex}   \sum_{\mbox{all possible} \ \{ A_\alpha \} \hfill \atop
\scriptstyle  \sum_{\alpha=1 }^P Q_\alpha= K+1 \  ; \  \left(  P \leq K+1  \right) }^{ \sum_{\alpha=1 }^P Q_\alpha A_\alpha  = 2L-K-2 }  \hspace{-2ex}\left[  \sum_{\beta =1}^P Q_\beta \left( A_\beta  + 3 \right)  \right] \nonumber  \\ 
& & \hspace{20ex} \times  \frac{1}{Q_1! \cdots  Q_P!} \frac{1}{\left( A_1 + 3 \right)^{Q_1} \cdots   \left( A_P + 3 \right)^{Q_P}}  \nonumber \\
& & = (-)^{K} \left( 2(L+K)+1 \right)!!     \nonumber  \\ 
& &  \hspace{6ex} \times  \sum_{\mbox{all possible} \ \{ A_\alpha \} \hfill \atop
\scriptstyle  \sum_{\alpha=1 }^P Q_\alpha= K+1 \  ; \  \left(  P \leq K+1  \right) }^{ \sum_{\alpha=1 }^P Q_\alpha A_\alpha  = 2L-K-2 }  \hspace{-2ex}\frac{1}{Q_1! \cdots  Q_P!}  \frac{1}{\left( A_1 + 3 \right)^{Q_1} \cdots   \left( A_P + 3 \right)^{Q_P}}   \ , \label{k=K} 
\end{eqnarray}
where use has been made of
\begin{eqnarray}
 \sum_{\beta =1}^P Q_\beta \left( A_\beta  + 3 \right) = 2(L+K)+1 \ ,
\end{eqnarray}
obtained from $ \sum_{\beta =1}^P Q_\beta A_\beta = 2L-K-2$ and $\sum_{\alpha=1 }^P Q_\alpha= K+1 $. The relation (\ref{Lemma}) has been proved.$\dbox$ 


Next we show the relation (\ref{no1/Ncontribution}) directly: to this end, prove a general formula
\begin{eqnarray}
& & \int_B^A dt g(t) {\mathrm e}^{-N f(t)} \stackrel{1/N}{=} \int_B^A dt  {\mathrm e}^{-N f(t)}\ .  \quad  g(t) = - f'(t) +1 \ ,  \label{DirecTarget}  
\end{eqnarray}
(In our case, $f(t)\equiv  t - \ln t, g(t) \equiv  1/t : B \equiv 0, A \equiv  \infty $.)  LHS of (\ref{DirecTarget}) becomes
\begin{eqnarray}
\mbox{LHS of (\ref{DirecTarget})} = \int_B^A dt  {\mathrm e}^{-N f(t)} -  \int_B^A dt f'(t) {\mathrm e}^{-N f(t)} \ ,
\end{eqnarray}
whose second term is integrable to yield
\begin{eqnarray}
\int_B^A dt f'(t) {\mathrm e}^{-N f(t)}= \int_{f(B)}^{f(A)} d f {\mathrm e}^{-Nf }= -\frac{1}{N}\left[ {\mathrm e}^{-Nf(A) } - {\mathrm e}^{-Nf(B) } \right] \ ,  \label{No1/Nterms}
\end{eqnarray}
implying no $1/N$ terms for any value of $Nf(A), Nf(B)$. We have proven (\ref{DirecTarget}). $\dbox$


In our case, from $A=\infty , B= 0$, that is, from $f(\infty ) = f(0)= \infty$,
\begin{eqnarray}
\int_0^\infty  dt {\mathrm e}^{-N \left( t- \ln t \right) } = \int_0^\infty  dt \ \frac{1}{t} \ {\mathrm e}^{-N \left( t- \ln t \right) }\ , \label{FormulaGammaCase}
\end{eqnarray}
which designates $\Gamma (N+1) = N\Gamma (N)$ by multiplying both sides by $N^N$.

\section{Calculation of ${\cal P}_0^{(\hat{i})} \sim {\cal P}_1^{(3)}$ and ${\cal L}_0^{(\hat{i})} \sim {\cal L}_1^{(3)} $ in (\ref{P'sExpansion}) and (\ref{L'sExpansion})\label{sec:appendixC}}
First expand $F^{(i)}(r); (r=1,2,3)$, (\ref{F(1)F(2) Def}) (\ref{F(3) Def}), such that
\begin{eqnarray} 
 F^{(i)}(1)  \equiv  \exp \left[ - \frac{N \left( \Omega_c^{(i)}- \omega  \right)^2 }{ 2 \lambda^2} \right]   = F^{(i)}_0 (1) \left[ 1 +  \frac{F^{(i)}_1 (1)}{N}  + \frac{F^{(i)}_2 (1)}{N^2}\right] \ , \label{F(1)} 
\end{eqnarray}
with
\begin{eqnarray} 
 & & \hspace{4ex}F^{(i)}_0 (1) \equiv \exp \left[ - \frac{N \left( \Omega_0^{(i)}- \omega  \right)^2 }{ 2 \lambda^2}  - \frac{\left( \Omega_0^{(i)}- \omega  \right) \Omega_1^{(i)}}{\lambda^2 }\right]  \ ,  \label{F_0(1)} 	\\
& & \hspace{-5ex} F^{(i)}_1 (1)  \equiv  -  \frac{ \left[ \left( \Omega_1^{(i)} \right)^2 + 2 \left( \Omega_0^{(i)}- \omega \right) \Omega_2^{(i)}  \right]}{2 \lambda^2}     \ , \  F^{(i)}_2 (1) \equiv  \frac{\left( F^{(i)}_1 (1) \right)^2 }{2} - \frac{\Omega_1^{(i)}\Omega_2^{(i)}}{\lambda^2}   \  .  \label{F_1(1) F_2(1)}	 
\end{eqnarray}
\begin{eqnarray}
   F^{(i)}(2)  \equiv \left( \Omega_c^{(i)} \right)^{N}  = F^{(i)}_0 (2) \left[ 1 +  \frac{F^{(i)}_1 (2)}{N}  + \frac{F^{(i)}_2 (2)}{N^2} \right]   \ , \label{F(2)} 
 \end{eqnarray}
 with
\begin{eqnarray}  
& & \hspace{-4ex}F^{(i)}_0 (2) \equiv  \left( \Omega _0^{(i)} \right)^{N}  \exp \left[ \frac{ \Omega _1^{(i)} }{ \Omega_0^{(i)} }  \right]  \ ,\   F^{(i)}_1 (2) \equiv \frac{  \Omega _2^{(i)}  }{ \Omega_0^{(i)} }   - \frac{1}{2}\left( \frac{ \Omega _1^{(i)} }{ \Omega_0^{(i)} }  \right)^2   \ , \label{F_0(2) F_1(2)} \\
& & F^{(i)}_2 (2) \equiv  \frac{\left(  F^{(i)}_1 (2) \right)^2 }{2}   - \frac{ \Omega _1^{(i)}   \Omega _2^{(i)} }{ \left( \Omega_0^{(i)}  \right)^2}   +  \frac{1}{3} \left( \frac{ \Omega _1^{(i)}  }{  \Omega_0^{(i)} } \right)^3 \ .\label{F_2(2)} 
\end{eqnarray}
(Note that there is no need for $F^{(3)}_2(r) ; (r=1,2)$.)
\begin{eqnarray} 
 F^{(i)}(3)  \equiv \left[ 1 + \left( \frac{\lambda}{\Omega_c^{(i)} }   \right)^2  \! \! \! + \frac{1}{N} \left( \frac{\lambda}{\Omega_c^{(i)} + \delta \omega}  \right) ^2  \right]^{-1/2} \hspace{-4ex}=  F^{(i)}_0 (3) \left[ 1 +  \frac{F^{(i)}_1 (3)}{N}  + \frac{F^{(i)}_2 (3)}{N^2} \right]   \ ,    \label{F(3)} 
\end{eqnarray}
with
\begin{eqnarray} 
& &  \hspace{0ex} F^{(i)}_0 (3) \equiv  \left\{ \hspace{-1ex}
 \begin{array}{l}
\displaystyle{  \frac{ \epsilon \! \left( \Omega_0^{(\hat{i})} \right) \Omega_0^{(\hat{i})}}{\sqrt{ \left( \Omega_0^{(\hat{i})} \right)^2 + \lambda^2  } }   }   \\
 \noalign{\vspace{1mm}}
\displaystyle{ \frac{ \epsilon \! \left( \Omega_1^{(3)} \right) \Omega_1^{(3)} }{\sqrt{N} \lambda }   }
                                    \end{array}
                                    \right.  \ ,   \label{F_0(3)}  \\
                                    & & F^{(i)}_1 (3) \equiv 
\left\{ \hspace{-1ex}
\begin{array}{l}
\displaystyle{  \frac{\lambda^2 }{\left( \Omega_0^{(\hat{i})} \right)^2 + \lambda^2} \left[ \frac{ \Omega_1^{(\hat{i})}}{ \Omega_0^{(\hat{i})} } - \frac{1}{2} \left( \frac{ \Omega_0^{(\hat{i})}}{ \Omega_0^{(\hat{i})} + \delta \omega } \right)^2  \right]       }  \\
\noalign{\vspace{1mm}}
\displaystyle{   \frac{ \Omega_2^{(3)} }{ \Omega_1^{(3)} }  -  \frac{ \left( \Omega_1^{(3)} \right)^2 }{2 \lambda^2 } - \frac{1}{2} \left(  \frac{ \Omega_1^{(3)} }{ \Omega_0^{(3)} }  \right)^2   }   
\end{array}      
\right.  .   \label{F_1(3)}   
\end{eqnarray}
\begin{eqnarray}  
 \hspace{-1ex}F^{(\hat{i})}_2 (3)  \equiv  \frac{ 3 \left( F^{(\hat{i})}_1 (3) \right)^2 }{2}   + \frac{\lambda^2 }{\left( \Omega_0^{(\hat{i})} \right)^2 + \lambda^2}\left[  \frac{ \Omega_2^{(\hat{i})}}{ \Omega_0^{(\hat{i})} } + \frac{\left( \Omega_0^{(\hat{i})} \right)^2 \Omega_1^{(\hat{i})}}{\left( \Omega_0^{(\hat{i})} + \delta \omega \right)^3 } - \frac{3}{2} \left(  \frac{ \Omega_1^{(\hat{i})}}{ \Omega_0^{(\hat{i})} } \right)^2  \right]    \ .  \label{F_2(3)}  
\end{eqnarray}
(Again no need for $F^{(3)}_2 (3)$.) Further $F^{(i)}(4)$,(\ref{F(4) Def}), for $i=1,2$ reads
\begin{eqnarray} 
 F^{(\hat{i})}(4)   = F^{(\hat{i})}_0 (4) \left[   1 +  \frac{ F^{(\hat{i})}_1 (4)}{N}   \right]    \ , \label{F(4)} 
\end{eqnarray}
with
\begin{eqnarray}
& & F^{(\hat{i})}_0 (4) \equiv \left( \Omega_0^{(\hat{i})}+ \delta \omega  \right)^2  \left(\left( \Omega_0^{(\hat{i})}  \right)^2+ \lambda^2   \right) \ ;  \label{F_0^hat(4)} \\
& & F^{(\hat{i})}_1 (4) \equiv  \frac{2 \Omega_1^{(\hat{i})}}{ \Omega_0^{(\hat{i})} +  \delta \omega } + \frac{2\Omega_0^{(\hat{i})}\Omega_1^{(\hat{i})} }{\left( \Omega_0^{(\hat{i})} \right)^2  +  \lambda^2 } + \frac{ \lambda^2 \left( \Omega_0^{(\hat{i})} \right)^2}{ \left(  \Omega_0^{(\hat{i})} + 
  \delta \omega  \right)^2  \left[  \left( \Omega_0^{(\hat{i})} \right)^2  +  \lambda^2  \right] } \ .  \label{F_1^hat(4)} 
\end{eqnarray}
Meanwhile for $i=3$
\begin{eqnarray}
NF^{(3)}(4) = F^{(3)}_0(4) \left[ 1 + \frac{ F^{(3)}_1(4) }{N} \right]    \ , \label{NF^3(4)}
\end{eqnarray}
with
\begin{eqnarray}
& & F^{(3)}_0(4) \equiv \lambda^2  \left(  \Omega_0^{(3)} \right)^2   \ , \label{F^3_0(4)} \\
& & F^{(3)}_1(4)       \equiv 2 \frac{ \Omega _1^{(3)} }{ \Omega_0^{(3)} }  + \left( \frac{ \Omega _1^{(3)} }{ \Omega_0^{(3)} }  \right)^2 +   \frac{ \left( \Omega_1^{(3)} \right)^2  }{ \lambda^2 }   \ . \label{F^3_1(4)}
\end{eqnarray}
$F^{(M;i)}(5); (M=3, \dots, 6)$, (\ref{F(5) Def}), for $i=1,2$ reads
\begin{eqnarray} 
 F^{(M;\hat{i})}(5)  = F^{(M;\hat{i})}_0 (5) \! \!\left[ \! 1 +  \frac{F^{(M;\hat{i})}_1 (5)}{N}    \right]  \ , \label{F^hat(5)} 
\end{eqnarray} 
with
\begin{eqnarray} 
 F^{(M;\hat{i})}_0 (5) \equiv   \left( \Omega_0^{(\hat{i})}+ \delta \omega  \right)^M  \ ,  \quad 
  F^{(M;\hat{i})}_1 (5) \equiv  \left( \frac{  \Omega_0^{(\hat{i})}  } { \Omega_0^{(\hat{i})} \! + \delta \omega} \right)^M \! \!  +  \! \frac{ M \Omega_1^{(\hat{i})} }{\Omega_0^{(\hat{i})} \! + \delta \omega}  \ .
\end{eqnarray}
Then for $i=3$
\begin{eqnarray}
NF^{(M;3)}(5)=  F^{(M;3)}_0 (5)\left[ 1 + \frac{ F^{(M;3)}_1 (5)}{N} \right]   \  ,  \quad  M=3, \dots, 6 \ , 
\end{eqnarray}
with
\begin{eqnarray} 
& & F^{(M;3)}_0 (5) \equiv \left(  \Omega_0^{(3)} \right)^M  \ ;   \qquad F^{(M;3)}_1 (5) \equiv M \frac{ \Omega _1^{(3)} }{ \Omega_0^{(3)} }  \ . 
\end{eqnarray}

Therfore the prefactor reads
\begin{eqnarray}
& & \hspace{-2ex}{\cal P}^{(\hat{i})}_0 \equiv F_0^{(\hat{i})}(1)F_0^{(\hat{i})}(2)F_0^{(\hat{i})}(3) \left( \Omega_0^{(\hat{i})}+ \delta \omega  \right)    \nonumber \\
& & \hspace{2ex}=  \exp \left[  - \frac{ N \left( \Omega_0^{(\hat{i})} - \omega  \right)^2 }{2 \lambda^2 } \right]    \left( \Omega_0^{(\hat{i})} \right)^{N+1}   \frac{ \epsilon \! \left( \Omega_0^{(\hat{i})} \right) }{\sqrt{ \left( \Omega_0^{(\hat{i})} \right)^2 + \lambda^2  } } \left( \Omega_0^{(\hat{i})} + \delta \omega \right)  , \label{P_0}
\end{eqnarray}
for $i=1,2$ and the leading term of $i=3$ is       
\begin{eqnarray}           
{\cal P}^{(3)}_1=    \exp \left[  - \frac{ N \left( \Omega_0^{(3)} - \omega  \right)^2 }{2 \lambda^2 } \right]    \left( \Omega_0^{(3)} \right)^N  \frac{ \epsilon \! \left( \Omega_1^{(3)} \right) \left( \Omega_1^{(3)} \right)^2  }{{\mathrm e} \sqrt{N} \lambda }     \ , \label{P_1(3)}
\end{eqnarray}
where use has been made of the relation
\begin{eqnarray}
  - \frac{ \left(  \Omega_0^{(i)} - \omega \right)  \Omega_1^{(i)} }{\lambda^2 }  +  \frac{\Omega_1^{(i)}}{\Omega_0^{(i)}} = \left\{
                                                                                                                            \begin{array}{cc}
                                                                                                                         0    &  i=1,2 	\\
  \noalign{\vspace{1ex}}                                                                                                                       -1    & i=3 
                                                                                                                            \end{array}
                                                                                                                            \right.   \ , 
\end{eqnarray}
in view of (\ref{Omega_cLeading}) and the gap equation (\ref{SaddlePoint Method1}) for $i=1,2$ or (\ref{Omega_1}) with (\ref{A&B}) for $i=3$. Then 
\begin{eqnarray}
& & {\cal P}^{(\hat{i})}_1 = \sum_{r=1}^3 F_1^{(\hat{i})}(r) +  \frac{\Omega_1^{(\hat{i})}}{ \Omega_0^{(\hat{i})} + \delta \omega }  \ ,   \label{P_1^hati} \\
& &  {\cal P}^{(3)}_2  =   \sum_{r=1}^3 F_1^{(3)}(r) +   \frac{\Omega_2^{(3)}}{\Omega_1^{(3)}}  \ , \label{P_2^3} 
\end{eqnarray}
and 
\begin{eqnarray}
{\cal P}^{(\hat{i})}_2 = \sum_{r=1}^3 F_2^{(\hat{i})}(r) + \frac{\Omega_2^{(\hat{i})}}{ \Omega_0^{(\hat{i})} + \delta \omega } + \sum_{r' > r=1}^3  F_1^{(\hat{i})}(r')  F_1^{(\hat{i})}(r) + \frac{\Omega_1^{(\hat{i})} }{\Omega_0^{(\hat{i})} + \delta \omega } \sum_{r=1}^3 F_1^{(\hat{i})}(r)   \ .  \label{P_2^hati}
\end{eqnarray}

Now calculate terms in the loop factor (\ref{L by F(4) & F(5)}) up to the $3$-loop approximation: 
\begin{eqnarray}
& & \hspace{-4ex}{\cal L}^{(i)}=   1   + \frac{\lambda^2}{N  F^{(i)}(4) }  \left[  - \frac{3\lambda^2}{4}   R^{(4;i)} + \frac{5 \lambda^4}{6} \left(  R^{(3;i)}  \right)^2  \right]   \nonumber \\
& &   \hspace{4ex} + \frac{\lambda^4}{\left( N   F^{(i)}(4) \right)^2  } \Bigg[  - \frac{5 \lambda^2}{2} R^{(6;i)}  + 7  \lambda^4 R^{(3;i)} R^{(5;i)}  +  \frac{105\lambda^4}{32} \left( R^{(4;i)} \right)^2   \nonumber \\
& &  \hspace{20ex}   \left. - \frac{105\lambda^6}{8}  \left( R^{(3;i)}  \right)^2 R^{(4;i)} + \frac{385\lambda^8}{72} \left(   R^{(3;i)} \right)^4  \right] \ , \label{LoopFactor}
\end{eqnarray}
where we have introduced
\begin{eqnarray}
 R^{(M;i)} \equiv \frac{ F^{(M;i)}(5)   }{  F^{(i)}(4)  } = R^{(M;i)}_0 \left[ 1 + \frac{R^{(M;i)}_1}{N} \right] +  O \! \left( \frac{1}{N^2} \right)  \ ,    \label{R Def} 
\end{eqnarray}
whose expansion coefficients are
\begin{eqnarray}
  R^{(M;i)}_0 \equiv \frac{ F^{(M;i)}_0(5)   }{  F^{(i)}_0(4)  } =\left\{
\begin{array}{c}
\displaystyle{ \frac{\left( \Omega_0^{(\hat{i})} + \delta \omega \right)^{M-2} }{\left( \Omega_0^{(\hat{i})} \right)^2 + \lambda^2 } } 	\\
 \noalign{\vspace{1mm}}                                                                    
 \displaystyle{ \frac{\left( \Omega_0^{(3)} \right)^{M-2} }{\lambda^2 } } 
  \end{array}
\right.  \ ,  \label{R_0^Mi} 
\end{eqnarray}
\begin{eqnarray}
& &  \hspace{0ex} R^{(M;i)}_1 \equiv  F^{(M;i)}_1(5) -  F^{(i)}_1 (4)  \nonumber \\
& & \hspace{-4ex} = \left\{ \hspace{-1ex}
\begin{array}{l}
\displaystyle{   \left( \!  \frac{ \Omega_0^{(\hat{i})} }{ \Omega_0^{(\hat{i})} \! \! +  \delta \omega } \!  \right)^M  \hspace{-2ex}+   \frac{(M-2) \Omega_1^{(\hat{i})}}{ \Omega_0^{(\hat{i})} \!  \! +  \delta \omega } -  \frac{2\Omega_0^{(\hat{i})}\Omega_1^{(\hat{i})} }{\left( \Omega_0^{(\hat{i})} \right)^2 \!  \! \!  +  \lambda^2 } -  \frac{ \lambda^2 \left( \Omega_0^{(\hat{i})} \right)^2}{ \left(  \Omega_0^{(\hat{i})} \! \! + 
  \delta \omega  \right)^2 \!   \left[ \!  \left( \! \Omega_0^{(\hat{i})} \!  \right)^2 \!  \! \!  +  \lambda^2 \!  \right] }    }  \\
 \noalign{\vspace{1mm}}
 \displaystyle{ (M-2) \frac{ \Omega _1^{(3)} }{ \Omega_0^{(3)} }  -  \left( \frac{ \Omega _1^{(3)} }{ \Omega_0^{(3)} }  \right)^2 -  \frac{ \left( \Omega_1^{(3)} \right)^2  }{ \lambda^2 }      }   
                                     \end{array}
                                    \right.   \ . \label{R_1^Mi}
\end{eqnarray}
With the aid of these, $i=1,2$ part is given by 
\begin{eqnarray}
& & \hspace{-6ex}{\cal L}^{(\hat{i})}_0=1  \  ; \label{L_0 i=1,2}   \\ 
& & \hspace{-6ex}{\cal L}^{(\hat{i})}_1 = \frac{\lambda^2 }{ \left(  \Omega_0^{(\hat{i})} \! \! + 
  \delta \omega  \right)^2 \!   \left[ \!  \left( \! \Omega_0^{(\hat{i})} \!  \right)^2 \!  \! \!  +  \lambda^2 \!  \right] } \left[  - \frac{3 \lambda^2 }{4} R^{(4;\hat{i})}_0  + \frac{5 \lambda^4 }{6} \left(  R^{(3;\hat{i})}_0  \right)^2  \right] \nonumber  \\
  & &   =    -  \frac{ 3 \lambda^4 }{4 \left[  \left( \! \Omega_0^{(\hat{i})} \!  \right)^2 \!  \! \!  +  \lambda^2 \!  \right]^2  } +   \frac{ 5 \lambda^6  }{6  \left[ \left( \! \Omega_0^{(\hat{i})} \!  \right)^2 \!  \! \!  +  \lambda^2 \!  \right]^3  }   \ , \label{L_1 i=1,2}  
\end{eqnarray}
\begin{eqnarray}
& & \hspace{-6ex}{\cal L}^{(\hat{i})}_2 =   \frac{\lambda^4}{  \left(  \Omega_0^{(\hat{i})} \! \! + 
  \delta \omega  \right)^4 \!   \left[ \!  \left( \! \Omega_0^{(\hat{i})} \!  \right)^2 \!  \! \!  +  \lambda^2 \!  \right]^2   }  \left[   - \frac{5 \lambda^2 }{2} R^{(6;\hat{i})}_0  + 7  \lambda^4 R^{(3;\hat{i})}_0 R^{(5;\hat{i})}_0  +  \frac{105 \lambda^4 }{32} \left( R^{(4;\hat{i})}_0 \right)^2  \right.  \nonumber \\
& &  \hspace{30ex} \left. - \frac{105 \lambda^6 }{8} \left( R^{(3;\hat{i})}_0  \right)^2 R^{(4;\hat{i})}_0 + \frac{385 \lambda^8 }{72} \left( R^{(3;\hat{i})}_0  \right)^4  \right] \nonumber \\
& &  \hspace{20ex}  - {\cal L}^{(\hat{i})}_1 F_1^{(\hat{i})}(4) - \frac{3 \lambda^4 }{4} \frac{   R^{(4;\hat{i})}_0  R^{(4;\hat{i})}_1   }{  F_0^{(\hat{i})}(4)  } +  \frac{5\lambda^6}{3} \frac{   \left(  R^{(3;\hat{i})}_0 \right)^2 R_1^{(3 ;\hat{i})} }{ F_0^{(\hat{i})}(4) }  \nonumber \\
& & =  -\frac{5}{2} \frac{\lambda^6}{   \left[ \!  \left( \! \Omega_0^{(\hat{i})} \!  \right)^2 \!  \! \!  +  \lambda^2 \!  \right]^3   } + \frac{329}{32} \frac{\lambda^8}{   \left[ \!  \left( \! \Omega_0^{(\hat{i})} \!  \right)^2 \!  \! \!  +  \lambda^2 \!  \right]^4   }  - \frac{105}{8} \frac{\lambda^{10}}{   \left[ \!  \left( \! \Omega_0^{(\hat{i})} \!  \right)^2 \!  \! \!  +  \lambda^2 \!  \right]^5   }  \nonumber \\
& & + \frac{385}{72} \frac{\lambda^{12}}{   \left[ \!  \left( \! \Omega_0^{(\hat{i})} \!  \right)^2 \!  \! \!  +  \lambda^2 \!  \right]^6   }  - {\cal L}^{(\hat{i})}_1 F_1^{(\hat{i})}(4) - \frac{3 \lambda^4 }{4} \frac{   R^{(4;\hat{i})}_0  R^{(4;\hat{i})}_1   }{  F_0^{(\hat{i})}(4)  } +  \frac{5\lambda^6}{3} \frac{   \left(  R^{(3;\hat{i})}_0 \right)^2 R_1^{(3 ;\hat{i})} }{ F_0^{(\hat{i})}(4) } \   . \label{L_2 i=1,2}
\end{eqnarray}

For $i=3$, in view of (\ref{NF^3(4)}), a naive $1/N$ expansion is broken down; all factors in front of the loop expansion, (\ref{L by F(4) & F(5)}), behave $O(1)$. Nevertheless we take up to the terms of $O \! \left( 1/\left\{  N F^{(3)}(4)  \right\} \right)$ under the 2-loop approximation, and of $O \! \left( 1/\left\{  N F^{(3)}(4)  \right\}^2 \right)$ under 3-loop so that
\begin{eqnarray}
 \hspace{-2ex}{\cal L}^{(3)}_0 \Bigg|_{2-\mbox{loop}} \hspace{-3ex} =    1   + \frac{\lambda^2}{F^{(3)}_0(4) }  \left[  - \frac{3 \lambda^2 }{4} R^{(4;3)}_0  + \frac{5 \lambda^4 }{6} \left(  R^{(3;3)}_0 \right)^2  \right] \! = \frac{13}{12}  \ , \label{L_0 i=3 2-loop}
\end{eqnarray}
with the use of (\ref{F^3_0(4)}) and (\ref{R_0^Mi}), and for $O\! \left( 1/N \right)$ terms we obtain 
\begin{eqnarray}
& &  \hspace{-3ex}{\cal L}^{(3)}_1 \Bigg|_{2-\mbox{loop}} \hspace{-3ex} =   \frac{\lambda^2}{F^{(3)}_0(4) } \left[  \left\{   - \frac{3 \lambda^2 }{4} R^{(4;3)}_0  + \frac{5 \lambda^4 }{6} \left(  R^{(3;3)}_0 \right)^2  \right\} \left( - F^{(3)}_1(4) \right) + \left\{   - \frac{3 \lambda^2 }{4} R^{(4;3)}_0 R^{(4;3)}_1 \right. \right.  \nonumber \\
& &  \hspace{18ex} \left.  \left. + \frac{5 \lambda^4 }{3} \left(  R^{(3;3)}_0 \right)^2 R^{(3;3)}_1 \right\}  \right] = - \frac{1}{12} F^{(3)}_1(4) -\frac{3}{4}  R^{(4;3)}_1+\frac{5}{3} R^{(3;3)}_1  \nonumber \\
& & \hspace{22ex} = -\left[ \left(\frac{ \Omega _1^{(3)} }{ \Omega_0^{(3)} }  \right)^2 +  \frac{ \left( \Omega_1^{(3)} \right)^2  }{ \lambda^2 } \right]  \ , \label{L_1 i=3 2-loop}
\end{eqnarray}
where use has been made of (\ref{F^3_0(4)}),(\ref{F^3_1(4)}), (\ref{R_0^Mi}), and (\ref{R_1^Mi}).

In a similar manner, we obtain the $O(1)$ term in $3$-loop, such that
\begin{eqnarray}
& &  \hspace{-2ex}{\cal L}^{(3)}_0  \Bigg|_{3-\mbox{loop}}  \hspace{-1ex} \! = {\cal L}^{(3)}_0 \Bigg|_{2-\mbox{loop}} \hspace{-3ex} +  \frac{\lambda^4}{ \left( F^{(3)}_0(4) \right)  ^2 } \left[   - \frac{5 \lambda^2 }{2} R^{(6;3)}_0   + 7  \lambda^4 R^{(3;3)}_0 R^{(5;3)}_0  +  \frac{105 \lambda^4 }{32} \left( R^{(4;3)}_0 \right)^2    \right.  \nonumber \\
& & \hspace{15ex}  \left. -  \frac{105 \lambda^6 }{8} \left( R^{(3;3)}_0  \right)^2 R^{(4;3)}_0   + \frac{385 \lambda^8 }{72} \left( R^{(3;3)}_0  \right)^4   \right]   = \frac{313}{288}    \ , \label{L^3_0 3-loop}
\end{eqnarray}
and for $O\! \left( 1/N \right)$ terms
\begin{eqnarray}
& &  \hspace{-4ex} {\cal L}^{(3)}_1\Bigg|_{3-\mbox{loop}} \hspace{-4ex}= {\cal L}^{(3)}_1 \Bigg|_{2-\mbox{loop}} \hspace{-1ex} + \frac{\lambda^4}{ \left( F^{(3)}_0(4) \right)  ^2 } \left[ \left\{   - \frac{5 \lambda^2 }{2} R^{(6;3)}_0   + 7  \lambda^4 R^{(3;3)}_0 R^{(5;3)}_0  +  \frac{105 \lambda^4 }{32} \left( R^{(4;3)}_0 \right)^2   \right.  \right.  \nonumber \\
& & \hspace{20ex}  \left. -  \frac{105 \lambda^6 }{8} \left( R^{(3;3)}_0  \right)^2 R^{(4;3)}_0   + \frac{385 \lambda^8 }{72} \left( R^{(3;3)}_0  \right)^4   \right\} \left( -2 F^{(3)}_1(4)  \right)   \nonumber \\ 
& & \hspace{4ex}  +   \left\{   - \frac{5 \lambda^2 }{2} R^{(6;3)}_0 R^{(6;3)}_1  + 7  \lambda^4 R^{(3;3)}_0 R^{(5;3)}_0 \left( R^{(3;3)}_1+ R^{(5;3)}_1 \right)   +  \frac{105 \lambda^4 }{16} \left( R^{(4;3)}_0 \right)^2  R^{(4;3)}_1  \right.  \nonumber \\
& & \hspace{8ex}  \left. \left.  -  \frac{105 \lambda^6 }{8} \left( R^{(3;3)}_0  \right)^2 R^{(4;3)}_0 \left( 2 R^{(3;3)}_1 +R^{(4;3)}_1 \right)  + \frac{385 \lambda^8 }{18} \left( R^{(3;3)}_0  \right)^4R^{(3;3)}_1 \right\}   \right]   \nonumber \\
& & \hspace{6ex}= - \frac{13}{144}   F_1^{(3)}(4) + \frac{137}{36} R_1^{(3;3)} - \frac{117}{16} R_1^{(4;3)} + 7 R_1^{(5;3)} - \frac{5}{2} R_1^{(6;3)}  \nonumber \\
& &  \hspace{6ex}=   - \frac{13}{12}\left[ \left( \frac{ \Omega _1^{(3)} }{ \Omega_0^{(3)} }  \right)^2 +  \frac{ \left( \Omega_1^{(3)} \right)^2  }{ \lambda^2 } \right]    \ .\label{L^3_1 Result} 
\end{eqnarray}

\end{document}